\documentclass{article}

\usepackage{times}
 
\usepackage{authblk}

\usepackage{graphicx}
\usepackage{epsfig}
\usepackage{verbatim}
\usepackage{amssymb}
\usepackage[final]{pdfpages}
\usepackage[labelfont=bf]{caption}
\usepackage{anysize}
\usepackage{amsfonts,amsmath}
\usepackage{bm}
\usepackage{multirow}
\usepackage[T1]{fontenc}
\usepackage{eufrak}
\usepackage{hyperref}
\usepackage{fancyhdr}
\usepackage{relsize}
\usepackage{xr} 
 \usepackage{booktabs,siunitx}
\usepackage{array,booktabs,ragged2e}

\providecommand{\be}{\begin{equation}}
\providecommand{\ee}{\end{equation}}
\providecommand{\bea}{\begin{eqnarray}}
\providecommand{\eea}{\end{eqnarray}}
\providecommand{\beas}{\begin{eqnarray*}}
\providecommand{\eeas}{\end{eqnarray*}}
\providecommand{\bal}{\begin{aligned}}
\providecommand{\eal}{\end{aligned}}
\providecommand{\bs}{{\bf s}}

\providecommand{\bM}{{\bm M}}
\providecommand{\bS}{{\bf S}}
\providecommand{\bH}{{\bm H}}
\providecommand{\Ham}{{\cal H}}

\hypersetup{colorlinks=true,
	linkcolor=blue,
	anchorcolor=black,
	citecolor=red,
	urlcolor=black
}

\begin{document}

\title{A statistical-inference approach to reconstruct inter-cellular interactions in cell-migration experiments} 
\author[1]{Elena Agliari}
\affil[1]{Dipartimento di Matematica, Sapienza Universit\`a di Roma, Rome, Italy}
\author[2]{Pablo J. S\'aez}
\affil[2]{UMR 144, Institut Curie, Paris, France}
\author[3]{Adriano Barra}
\affil[3]{Dipartimento di Matematica \& Fisica `Ennio De Giorgi', Universit\`a del Salento, Lecce, Italy}
\author[2]{Matthieu Piel}
\author[2]{Pablo Vargas}
\author[4,5]{Michele Castellana \footnote{Corresponding author. E-mail: michele.castellana@curie.fr.}}
\affil[4]{Laboratoire Physico-Chimie Curie, Institut Curie, PSL Research University, CNRS UMR 168, Paris, France}
\affil[5]{Sorbonne Universit\'es, UPMC Univ. Paris 06, Paris, France}

\maketitle 

\begin{abstract}
Migration of cells can be characterized by two, prototypical types of motion: individual and collective migration.  
We propose a statistical-inference approach designed to detect the presence of cell-cell interactions that give rise to collective behaviors in cell-motility experiments. Such inference method has been first successfully tested on synthetic motional data, and then applied to two experiments. 
In the first experiment, cell migrate in a wound-healing model: when applied to this experiment, the inference method predicts the existence of cell-cell interactions, correctly mirroring the strong intercellular contacts which are present in the experiment.
In the second experiment, dendritic cells migrate in a chemokine gradient. Our inference analysis does not provide evidence for interactions, indicating that cells migrate by sensing independently the chemokine source. According to this prediction, we speculate that mature dendritic cells disregard inter-cellular signals that could otherwise delay their arrival to lymph vessels. 
\end{abstract}

\section{Introduction}\label{intro}

Cell migration is a dynamic process, which may be characterized by two prototypical kinds of motion: a collective motion in which cells communicate with each other, e.g., by means of biochemical or mechanical signals, and an individual motion, where each cell migrates independently \cite{Helvert2018,Clark2015}.

Among the notable examples of cell migration in pathological contexts are cancer cells during metastasis,  where collective migration generally results from physical contacts between cells, which adhere to each other via specific adhesion molecules. Despite the fact that both individual and collective motions have been observed, for cancer cells collective migration is believed to result in more efficient metastatic spreading \cite{carmona2008contact,mayor2010keeping}. 

On the other hand, immune cells do not exhibit cell-to-cell adhesion during migration: for these cells, the existence of physical cell-to-cell communication is induced by specific signaling pathways \cite{Saez2014},
and restricted to slow migratory phases. Importantly, such direct cell-to-cell communication allows immune cells to share pathogenic information or to coordinate the arrival of other cells \cite{Kreisel2010, Majumdar2019}, while it does not result in a collective motion. As a consequence, in the context of fast migration if immune cells, such cells have been suspected to migrate as single cells.
 
However, immune cells respond to a large variety of biochemical signals. They release cytokines, chemokines and small molecules that control cell migration, and may send signals to adjacent cells without the need of physical interactions \cite{Pablo2017, Majumdar2019}.
This paracrine signaling might constitute an alternative mechanism to achieve cell-to-cell communication which, as shown in other cellular systems \cite{Kriebel2018}, may ultimately lead to cellular coordination. As a result, the existence of these signaling mechanisms raise the question of whether immune cells may ultimately migrate `collectively', regardless of physical interactions. 

Among the difficulties in answering this question is the fact that immune cells are often guided by extracellular signals produced by tissues \cite{Tiberio2018}: as a result, they may be exchanging biochemical signals with each other, but still appear to migrate individually because they all move towards the {\em same}  signal source.  
In addition, the number of potential molecules that could be responsible for cell-to-cell signaling is so large that molecular-perturbation approaches could not rule out the existence of cell-to-cell communication through some of these molecules. Moreover, molecular-perturbation approaches rely on an a priori knowledge of the signaling involved, and/or the development of new experimental tools which are, in general, both time consuming and expensive.\\

In this study, we propose a  statistical-inference method  to overcome the issues mentioned above. Unlike the classical molecular-perturbation approaches, our procedure is {\em statistically} rather than {\em biologically} driven, and does not rely on any a priori knowledge of the biochemical interactions among the migrating cells. Instead, our method leverages the statistical information stored in empirical observations on cell motility, e.g., statistics of cells' speeds and directions of motion \cite{Agliari2016NSR,Gigli2019}. We designed this statistical-inference method so as to take account of a variety of experimental sources of error that are specific to cell-motility experiments, e.g., limited tracking resolution, missing trajectories and tracking anomalies, and thus efficiently exploit the information contained in the cell-tracking data. 
Given a set of data for a cell-migration experiment, the resulting statistically inferred model allows one to single out cell-cell interactions. While the method cannot assess the nature of these interactions, e.g., the signaling protein that is responsible for it, it  makes a clear prediction on the existence, or absence, of such interaction.  In particular, the model allows us to tell apart a population of independent cells, which may appear to migrate collectively only because they  all follow the same cue, from a population that migrates in a truly collective manner. 

We first test the inference method on two benchmark data sets, i.e., synthetic cell trajectories generated from a mean-field model of interacting spins and  from a non-mean-field model of self-propelled particles.  In both cases, the method correctly reconstructs the microscopic features of the models, e.g., the spin-spin couplings, and the strength of the external signal to which the self-propelled particles are subject.

Next, we applied the inference framework to two experiments: In the first, cells migrate towards a `wound', and physically interact with each other via cell-cell adhesion resulting from cellular crowding. The inference method predicts the existence of cell-cell interactions, and this prediction correctly reflects the presence of intercellular contacts in the experiment, thus yielding a proof of concept for our inference framework.  

We then applied the inference method to a  population of dendritic cells migrating in a chemokine gradient. Our inference analysis does not provide evidence for the existence of a cell-cell interactions, indicating that dendritic cells migrate independently, and sense only the chemokine source. This prediction is supported by an exhaustive analysis of the raw motional data, which indicates the absence of cell-cell correlations.  

Finally, we present the biological implications of our results, and discuss how the predicted absence of cell-cell interactions may constitute a strategy to efficiently trigger the immune response. 
 
\section{Results}\label{res}

\subsection{Maximum-entropy methods}\label{mei}

In what follows, we will describe the statistical-inference method that we designed for the analysis of motional features of cell-tracking experiments. Given a  data set ${\bf X} = \{x_1, \cdots, x_T\}$ composed of multiple observations of a quantity $x$, maximum-entropy (ME) models provide a fundamental principle to model and reconstruct the probability distribution $P(x)$ from a limited number of empirical observations, which would be too small to reconstruct such distribution directly from the data. Specifically, given a set of features $f_1(x), f_2(x), \cdots$ related to the observable $x$, their  experimental and model estimates are, respectively,
\bea\label{av_eq}
\langle f_i(x) \rangle_{\rm ex} &=& \frac{1}{T} \sum_{t=1}^T f_i(x_t),\\
\langle f_i(x) \rangle_{P} &=&   \displaystyle\int dx P(x) f_i(x).
\eea
The ME method then constructs $P$ as the least-structured probability distribution that matches the experimental averages above.
Given that  the amount of `structure' in $P$ is quantified by the entropy \cite{shannon1948a}
\be\label{eqS}
S[P] = - \int dx P(x) \log P(x),
\ee
i.e., the higher $S$, the less structured $P$, the ME model is formulated in terms of the following constrained optimization problem
\bea
\label{eq_ent}
  \underset{P}{\max} \, S[P] &&\\ \nonumber
\textrm{subject to}&&\\
\langle f_i(x) \rangle_P   =  \langle f_i(x) \rangle_{\rm ex},&& \label{eq_const}\\
\int dx P(x)  =   1.&& \label{eq_norm}
\eea

ME models have attracted growing interest in the past few years, and have been used for a wide variety of domains and  of biological systems. Notable examples are the inference of motional order in flocks of birds \cite{bialek2012statistical}, collective behavior in networks of neurons \cite{schneidman2006weak,shlens2006the},  interaction structures resulting from amino-acid sequences in protein families \cite{seno2008maximum,weigt2009identification} and interaction structure of genetic networks \cite{lezon2006using}.\\

\begin{figure}
\begin{center}
\includegraphics[scale=0.7]{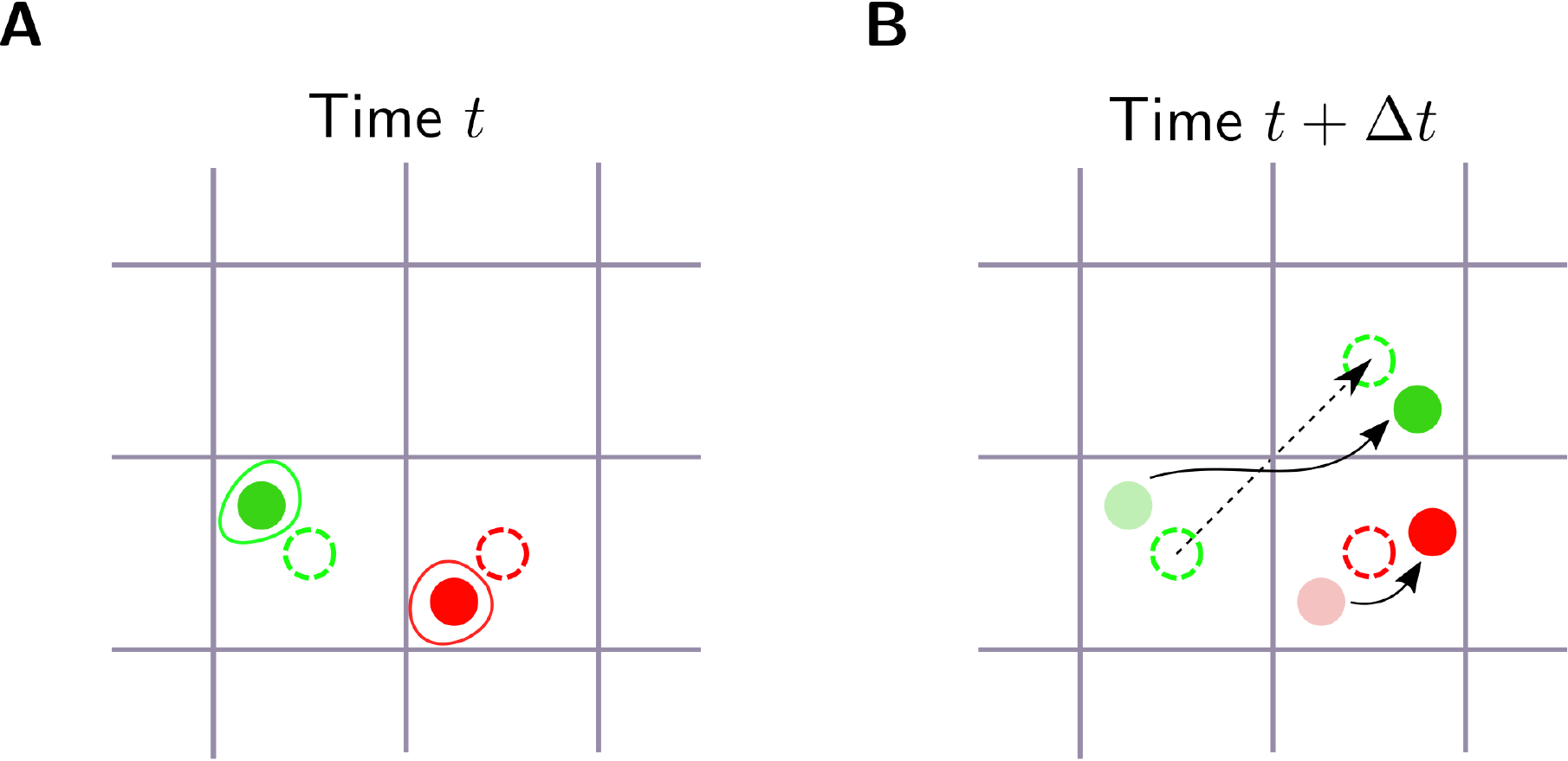}
\caption{\label{fig:needKKT}
\textbf{Motivation for maximum-entropy models with bound constraints in cell-tracking experiments}. ({\textbf A}) Two cells (whose boundaries are the red and green closed curves) and their centers of mass (red and green disks) are tracked through a grid of pixels (gray). Each center is assigned a nominal position, i.e., the center of the pixel where it is located (dashed red and green circles, respectively). (\textbf{B}) At a subsequent time, the green cell has moved to a neighboring pixel (curved black arrow). The  nominal position of the cell has changed, and its displacement is  the vector difference between the nominal position in (\textbf{B}) and in ({\textbf A}) (dashed black arrow), resulting in a well-defined direction of motion. Because the red cell has moved within the pixel, its nominal position is the same as in ({\textbf A}),  its nominal displacement is null, and its direction of motion is not defined, thus leading to an uncertainty in all physical observables which involve directions of motion. }
\end{center}
\end{figure}

In the specific problem under consideration in this study, the positions  of $N$ moving cells are imaged and tracked in time with a camera on a pixel grid, see Fig. \ref{fig:needKKT}.
Given that the precision with which the cell position is determined cannot exceed the pixel size, at any instant of time $t$ every cell, labeled by index $i$, is assigned a nominal position ${\bf r}_i(t)$, which coincides with the center of the pixel---see Supplementary Material, Section \ref{sec_delta} for details.  The nominal cell positions ${\bf r}_i(t)$ and ${\bf r}_i(t+\Delta t)$ at time $t$ and at a subsequent observation $t+\Delta t$, respectively, yield the  velocity ${\bf v}_i(t) = [{\bf r}_i(t+\Delta t) - {\bf r}_i(t)]/\Delta t$,  where we use boldface for vector quantities.
We compute the direction of motion, i.e., the normalized velocity, of each cell at time $t$
\be\label{eq_s}
\bs_i(t) = \frac{{\bf v}_i(t)}{|{{\bf v}_i(t)}|},
\ee
and obtain the full set of motional directions of the population, $\bS_t = \{\bs_1(t),\cdots, \bs_N(t)\} $, which we regard as the empirical observations for the ME problem, i.e., $x_t = \bS_t$ \cite{bialek2012statistical}. We select as features the average pairwise correlation and polarization
\bea\label{eq_c}
f_1(x) & = &  \frac{1}{N_p} \sum_{i<j=1}^N \bs_i \cdot \bs_j \equiv C(\bS),\\ \label{eq_m}
{\bm f}_2(x) & = &\frac{1}{N} \sum_{i=1}^N \bs_i  \equiv  \bM(\bS),
\eea
respectively, where $N_p \equiv N(N-1)/2$ is the number of cell pairs, and obtain the ME distribution by solving the optimization problem  (\ref{eq_ent})-(\ref{eq_norm}), see Section \ref{Sec:Methods-1} for details. If the resulting ME distribution factors out as the product of distribution of independent directions of motion, we obtain that cells behave independently; if not, we conclude that there exists an interaction between cells. Importantly,  the feature choice (\ref{eq_c}) sets the type of interaction that the ME method will probe. For instance,  the experimental average $\langle \rangle_{\rm ex}$ of Eq. (\ref{eq_c}) involves directions of motion $\bs_i(t) \cdot \bs_j(t)$ of cells $i$ and $j$ evaluated at the same instant of time $t$: it follows that the resulting cell-cell interaction inferred by the ME model will necessarily be an instantaneous one, i.e., its propagation time is significantly shorter than all other time scales \cite{schneidman2006weak}. Alternative types of interactions could be probed by choosing other features, e.g., by introducing a lag between times at which $\bs_i$ and $\bs_j$ are evaluated.   \\

Despite its wide use in a variety of systems \cite{bialek2012statistical,schneidman2006weak,weigt2009identification}, the ME method above may suffer from a fundamental limitation when applied to data affected by strong uncertainties. In fact, if the empirical data contains significant errors or a limited amount of information, complete satisfaction of the equality constraint is known to be too strict a criterion, and may lead to data overfitting \cite{kazama2005maximum, chen2000survey}. 
A prototypical example of this  issue comes from ME models for language modeling \cite{chen2000survey}, where the observations $x = (w, w')$ are pairs of consecutive words, $w$ and $w'$, in a corpus of text, which constitutes the data set $X$. Given two words, e.g., `$\rm saint$' and `$\rm George$', we consider as features the frequency with which `$\rm George$'  occurs in the text $f_1(w, w') = \mathbb I(w' = {\rm George})$,  and the frequency of the bigram `$\rm saint\;  George$', $f_2(w, w') = \mathbb I(w = {\rm saint},w' = {\rm George})$, where the indicator function $\mathbb I$ is one if both the conditions in its argument are satisfied, and zero otherwise.
Given a limited amount of empirical information, e.g., a short corpus of text where  the word `$\rm George$' occurs only after `$\rm saint$', if we impose these constraints in their equality form (\ref{eq_const}), it is straightforward to show that $P(w, \, {\rm George}) = 0$ if $w \neq \rm `saint'$. These zero-frequency events in the ME model not only may cause numerical instability in ME estimation \cite{kazama2005maximum}, but may also result in poor performance of the ME model in  a variety of applications, e.g., text recognition, where any word pair $(w, \rm George)$ in which  $w \neq `\rm saint'$ would not be recognized as a bigram.\\

The effect of data uncertainties may be even more dramatic in cell-tracking experiments. As shown in Fig. \ref{fig:needKKT}, if the cell motion is slow compared to the rate at which the observations are collected, the nominal  position ${\bf r}(t+\Delta t)$  at time $t+\Delta t$ may coincide with ${\bf r}(t)$, and the direction of motion (\ref{eq_s}) is not defined. As a result, any empirical average which involves the directions of motion, e.g., the polarization (\ref{eq_m}), will be affected by an error and will not be uniquely determined from the data, thus making the classical ME formulation pointless.

A potential workaround for this issue would be to measure the  positions at intervals  larger than $\Delta t$, in such a way that two subsequent measurements of the cell position ${\bf r}(t)$ lie in different bins for all cells and  times: the resulting empirical averages could then be analyzed with the classical ME formulation, see Section \ref{Sec:Methods-1}. However, this strategy would throw away a large number of the original measurements ${\bf r}(t), \, {\bf r}(t+\Delta t), \, \cdots$, 
and thus use only a fraction of the experimental data available. In what follows, we discuss a ME formulation with bound constraints (MEb)  \cite{kazama2005maximum, chen2000survey}, that efficiently exploits all the information contained in the experimental data, see Section \ref{meb} for details. In addition to the uncertainty above on the directions of motion, this MEb method may be used to handle a variety of other experimental sources of error, such as missing tracks, tracking anomalies, and others.\\

Let us suppose that, because of the experimental uncertainty described in Fig. \ref{fig:needKKT}, the tracking data does not provide  a precise value for the empirical averages, but a confidence interval in which these averages lie: Namely, if we let each unknown direction of motion $\bs_i$ vary between $0$ and $2 \pi$, then $\langle C \rangle_{\rm ex}$ and $\langle {\bm M} \rangle_{\rm ex}$ will fluctuate between a lower and an upper bound, which define a confidence interval.
As a result, in the MEb approach  we introduce explicitly such confidence interval by smoothening the equality constraints in ME model (\ref{eq_ent})-(\ref{eq_norm}) \cite{chen2000survey}: the equality constraint (\ref{eq_const}) are replaced by
\be\label{const_ineq}
\langle f_i(x) \rangle_{\rm ex}^{\rm min} \leq \langle f_i(x) \rangle_P   \leq \langle f_i(x) \rangle_{\rm ex}^{\rm max},
\ee
where  $\langle f_i(x) \rangle_{\rm ex}^{\rm min, max}$ are the lower and upper bounds for the empirical average of feature $f_i$, respectively \cite{kazama2005maximum}.

\subsection{Statistical-inference analysis}\label{meb_synthetic}

In what follows, we will describe the main features of the MEb method, and refer to Section \ref{meb} for details.
The joint distribution of velocities resulting from the MEb construction has the shape of a Boltzmann distribution
\bea \label{P}
P(\bS) = \frac{1}{Z} e^{-\Ham(\bS)},
\eea
with Hamiltonian
\be\label{eq_H}
\Ham(\bS) = -N[ J C(\bS) + {\bH}\cdot \bM(\bS)],
\ee
where $J$ reflects the ``interaction'' between cell velocities, $\bH$, the ``external field'', represents the overall tendency of the cells to flow in one particular spatial direction, and the partition function $Z$ ensures that $P$ is normalized. The Hamiltonian (\ref{eq_H}) is the one of the mean-field XY model in statistical mechanics \cite{huang1987statistical}, see Section \ref{calibro-XY} for details. The solution of the MEb problem is determined by a set of equality and inequality conditions---also denoted by bound constraints---known as the Karush-Kuhn-Tucker (KKT) conditions \cite{karush1939minima,kuhn1951nonlinear}. Given that each of the three parameters that appear in $P$, i.e., $J$ and the two components of $\bH$, can be either positive, negative, or zero, we obtain a set of candidate MEb solutions, where each solution corresponds to a sign configuration of the parameters above. The MEb solution is then given by the solution with the largest entropy, and that satisfies all equality and inequality constraints---see Sections \ref{meb_cancer} and \ref{meb_dendritic} for details.\\

We tested the MEb  on extensive synthetic data generated with  a mean-field XY model and a self-propelled model (SP) of particles, finding overall a very good agreement, see Section \ref{Sec:Methods-calibration} for details.
Given the satisfactory results of the MEb on these benchmark data sets, in what follows we will present the experimental data of our study, discuss their statistical features, and analyze them with the MEb. 

\subsection{Experiments}\label{sec:C}

\begin{figure}
\begin{center}
\includegraphics[scale=.18]{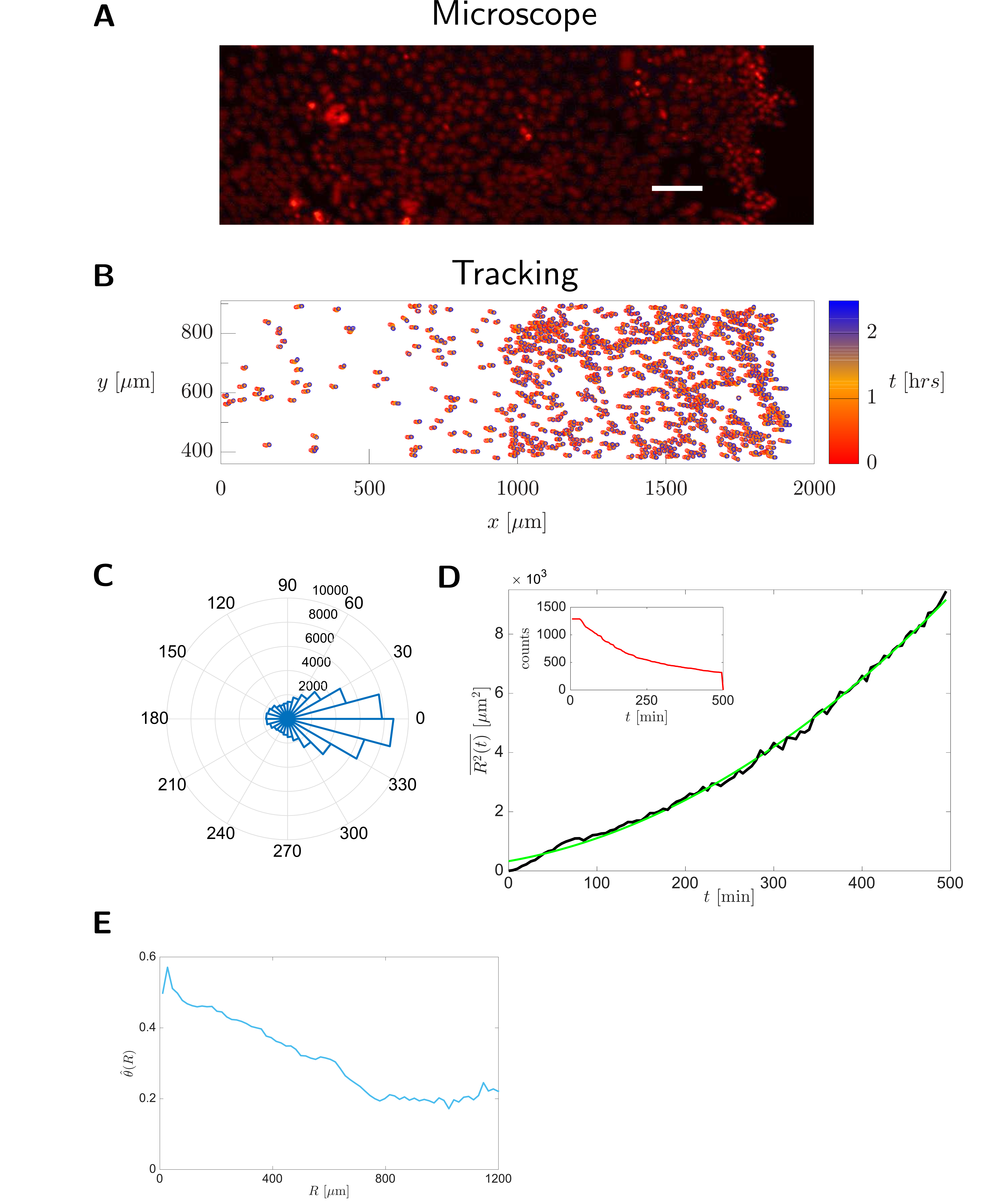}
\caption{\label{figA} 
\textbf{Wound-healing experiment}. ({\textbf A}) Fluorescence image showing the nuclei (red, H2B-mCherry) of HeLa cells migrating towards a wound located at the right edge of the image (scale bar $100\, \mu \rm m$). (\textbf{B}) Tracked cell trajectories: the instantaneous position of each cell is marked with a circle, an the respective time is specified by the color code. Only a few representative cells are shown for clarity.
(\textbf{C}) Polar histograms for the angle $\theta$ and (\textbf{D}) mean-squared displacement $\overline{R^2(t)}$ versus time show that the motion is affected by a strong bias which yields  a mean-squared displacement growing quadratically in time: Experimental data are shown in dark color, and the best fit $y = p_1 + p_2 t + p_3 t^2$ with $p_1 = 33, p_2 =0.5, p_3 =0.01$ is shown in bright color.
The inset in (\textbf{D}) shows the number of tracks recorded at each instant of time in order to figure out the time-window statistically significant. 
 (\textbf{E}) Angle pairwise correlation $\hat{\theta}(R)$ for the wound-healing experiment. }
\end{center}
\end{figure}

\begin{figure}
\begin{center}
\includegraphics[scale=0.16]{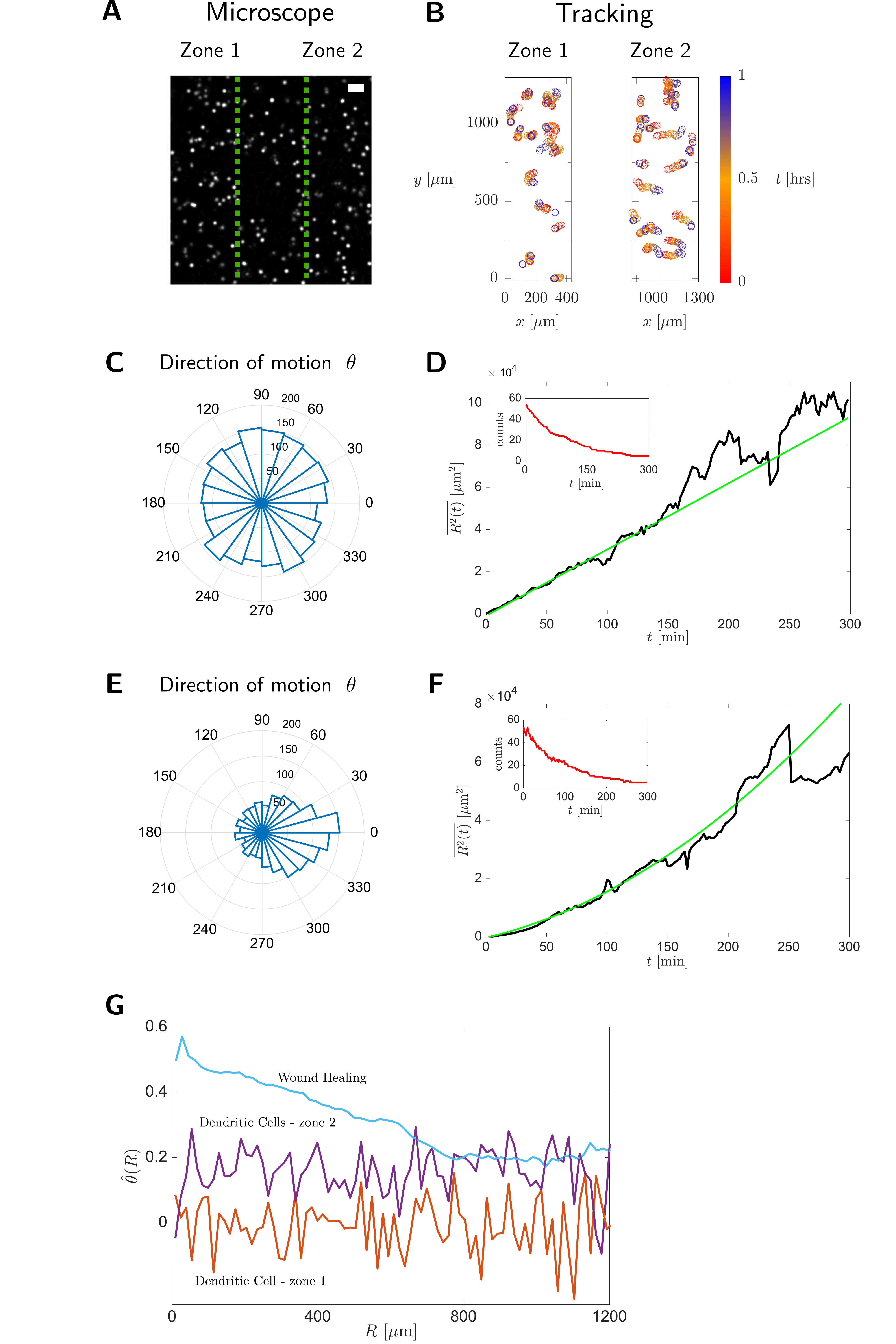}
\caption{\label{figB}
\textbf{Dendritic-cell experiment}. 
(\textbf{A}) Microscope image of cells showing chemokine-poor and chemokine-rich regions, i.e., zones 1 and 2, respectively, separated by green dashed lines  (scale bar $100\, \mu \rm m$). (\textbf{B}) Tracked cell trajectories in zone 1 and zone 2: the instantaneous position of each cell is marked with a circle, an the respective time is specified by the color code. Only a few representative cells are shown for clarity. Cells in zone 1 (\textbf{C} and \textbf{D}) move isotropically and diffusively (best fit $y = p_1 + p_2 t$ with $p_1 = -90, p_2 =31$), while cells in zone 2 (\textbf{E} and \textbf{F}) feel a drift and move ballistically ($y = p_1 + p_2 t + p_3 t^2$ with $p_1 =-306, p_2 =99, p_3 =0.6$). The insets in (\textbf{D}) and (\textbf{F}) show the number of tracks recorded at each time.  (\textbf{G}) Angle pairwise correlation $\hat{\theta}(R)$ for the three experimental instances above.
}
\end{center}
\end{figure}

In the wound-healing experiment, a population of human cancerous epithelial cells migrates in a planar device in which we realized a wound, see Fig. \ref{figA}  and Section \ref{sec:C2} for details. In the dendritic-cell experiment, cells move in a spatially varying concentration of chemokines built along the horizontal axis of the device, see Fig. \ref{figB} and Section \ref{sec:C1} for details.

\subsubsection{Analysis of cellular trajectories}\label{sec:C3}

For both experiments, the data consists of the nominal coordinates ${\bf r}_i(t)$, $t=1,...,T_i$, where the length $T_i$ of track $i$ is cell dependent because the tracks may lie in the observation window for different periods, and the time lapse between two measurements is $\Delta t = 5 \, \rm min$ and $\Delta t = 2 \, \rm min$, for the wound-healing and dendritic-cell experiment, respectively, see Sections \ref{sec:C2} and \ref{sec:C1} for details. Therefore, tracks are regarded as discrete-time random walks, where the $i$-th walker at time $t$ takes a step $\Delta {\bf r}_i(t) = {\bf r}_i(t+\Delta t) - {\bf r}_i(t)$, with length $\Delta r_i(t) = |\Delta {\bf r}_i(t)|$ and velocity ${\bf v}_i(t)$.
In addition to the Cartesian coordinate system above, we  describe the motion in a polar system where, at the $t$-th time step, the $i$-th track performs a step of length $\Delta r_i(t)$ in the direction described by the angle
$\theta_i(t)$ with respect to the horizontal axis.

In the remainder of this Section, we introduce the fundamental motional features of the cells tracked in the two experiments, in view of the statistical analysis of Sections \ref{meb_cancer} and \ref{meb_dendritic}.
First, we  introduce the  displacement of cell $i$, i.e., $\mathbf{R}_i(t) = \mathbf{r}_i(t) - \mathbf{r}_i(0)$ and its square average $\overline{R^2(t)}$ over all available tracks:
\begin{equation}
\overline{R^2(t)} = \frac{1}{N_t} \sum_{i=1}^{N} \mathbb{I}(T_i \geq t) |\mathbf{R}_i(t)|^2,
\end{equation}
where the normalization $N_t$ is the number of cells whose tracks have length larger or equal than $t$, i.e.,  $N_t = \sum_{i=1}^N \mathbb{I}(T_i \geq t)$.
Also, we introduce the pairwise correlation of the angle $\theta$ versus the distance $R$ between cells, i.e., 
\begin{equation}
\hat{\theta}(R) =  \frac{1}{N_R} \sum_{i,j=1}^N \sum_{t=1}^{\min (T_i,T_j)} \mathbb{I}(|\mathbf{r}_i(t)-\mathbf{r}_j(t)|=R)  \theta_i(t)\theta_j(t),
\end{equation}
where the normalization $N_R$ is the total number of pairs at distance $R$, namely $N_R = \sum_{i<j=1}^N \mathbb{I}(|\mathbf{r}_i(t)-\mathbf{r}_j(t)|=R)$.\\

In what follows, we will discuss shortly the results of the analysis of motional data, and refer to Section \ref{sec:method-statisticaltreatment} for details. 

In agreement with previous observations, in the wound-healing experiment the trajectories showed a strong bias along the horizontal direction evidenced by a sharply-peaked polar histogram for the angle $\theta$ (Fig. \ref{figA}C) and a ballistic-like mean-squared displacement $\overline{R^2(t)} \propto  t^2$ (Fig. \ref{figA}D). In addition, the angle pairwise correlation among cells are non-null, and it decays with respect to cell-cell distance (Fig. \ref{figA}E).

In the dendritic-cell experiment, we found different behaviors depending on the proximity to the chemokine-rich region \cite{Pablo2018}. In chemokine-free region (zone 1), cells exhibit an isotropic random walk with an uniform polar histogram for the angle of migration $\theta$ (Fig. \ref{figB}C), and a diffusive-like mean-squared displacement $\overline{R^2(t)} \propto  t$ (Fig. \ref{figB}D). On the other hand, in the region where there is a chemokine gradient (zone 2), cells perform a biased walk with a peaked polar histogram for the angle $\theta$ (Fig. \ref{figB}E), and a ballistic-like mean-squared displacement $\overline{R^2(t)} \propto  t^2$ (Fig. \ref{figB}F). Moreover, Fig. \ref{figB}G shows that angle pairwise correlation is absent in zone 1 and non-null in zone 2, and that in both cases no statistically significant dependence on $R$ is found.
Importantly, Fig. \ref{figB}G shows that the correlation decay with distance is peculiar to the dendritic-cell experiment, and it suggests that diverse migratory mechanisms may be at work in the two experiments, pointing to the existence of a collective behavior in the wound-healing experiment. 
In what follows, we will leverage the MEb method in an effort to unveil the presence of these mechanisms, and find out whether cell motion is simply gradient driven, or whether a collective migration is also at play.

\subsubsection{Statistical-inference analysis: wound-healing experiment}\label{meb_cancer}

\begin{figure}
\begin{center}
\includegraphics[scale=1.9]{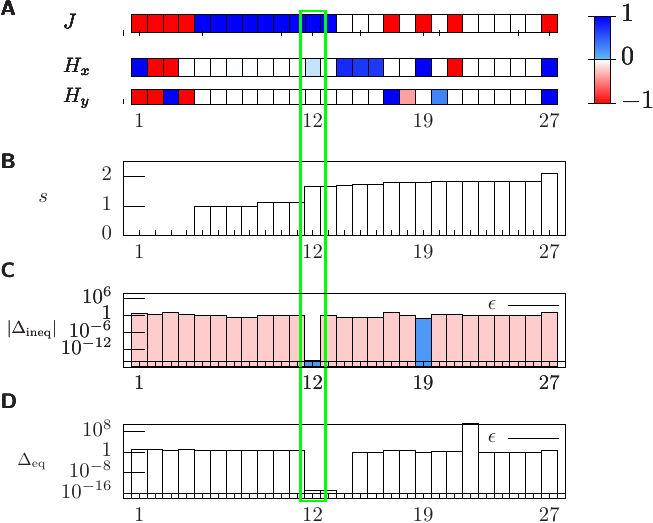}
\caption{
\textbf{Analysis of the wound-healing experiment via maximum-entropy method with bound constraints}.
({\textbf A}) Full set of solutions of the Karush-Kuhn-Tucker (KKT) conditions,  in order of increasing entropy from left to right. Each solution is labeled with an integer shown on the abscissa, and the corresponding value of $J$ (top), $H_x$ (middle) and $H_y$ (bottom) are represented with the color in the box. (\textbf{B}) Entropy per cell for each  solution. (\textbf{C}) Modulus of  the relative residual $\Delta_{\textrm{ineq}}$ of the inequality KKT conditions shown for each  solution, where residuals that are positive and negative are marked in red and blue, respectively.  (\textbf{D})  Modulus of relative residual $\Delta_{\textrm{eq}}$ of the equality KKT conditions, shown for each  solution. The numerical precision used in the calculation, $\epsilon$, is  marked in (\textbf{C}) and (\textbf{D}).
The maximum-entropy solution is marked with a green rectangle, and in  (\textbf{B}) unphysical solutions with  imaginary entropy are not shown.}
\label{fig4}
\end{center}
\end{figure}

In order to analyze the cell trajectories with the MEb, we estimated the uncertainty on cell positions resulting from the finite pixel size, see Fig. \ref{fig:needKKT} and Section \ref{sec_delta} for details, incorporated this uncertainty in the lower and upper bounds of empirical averages as described in Section \ref{mei}, and used the MEb with these bounds.
 
First, we observe that the spread of the correlation average due to the experimental error, i.e., $(\langle C\rangle_{\rm ex}^{\rm max}-\langle C\rangle_{\rm ex}^{\rm min})/(\langle C\rangle_{\rm ex}^{\rm max}+\langle C\rangle_{\rm ex}^{\rm min})$, can be as large as $\sim 50 \%$, see Table \ref{Table1}. Such a large spread indicates that it would be pointless to rely on the empirical averages only by using the classical ME method \cite{bialek2012statistical}, thus demonstrating the  need for a MEb formulation. The MEb solution is depicted in Fig. \ref{fig4}: A visual inspection of the full set of candidate solutions shows that there is a unique solution that has the largest entropy per cell $s = S/N$, and which satisfies all constraints within numerical precision.
As shown in Fig. \ref{fig4}, the results of the MEb analysis are $J = 1.1$, $\bH = (0.018,0)$. Because $J$ and $\bH$ are multiplied by $O(N)$ terms in the Hamiltonian (\ref{eq_H}), these parameters should be considered to be significantly different from zero as long as they are of order unity: it follows that the MEb solution yields an interaction parameter $J$ significantly different from zero, thus indicating the presence of a collective behavior in the cell migration of Fig. \ref{figA}. In addition, the small value of the $x$ component of the external field may be ascribed to the attractive effect of the wound located on the right edge of the observation window, see Fig. \ref{figA}A. Finally, the null $y$ component of the external field reflects the spatial homogeneity of the experiment with respect to the vertical direction.\\

It is important to point out that the result $J \neq0$ is consistent with the analysis of motional data of Sections \ref{sec:C3} and \ref{sec:SecondExperiment_RW}. In fact, that analysis shows that the empirical distribution of the cell velocities is  exponential in the $y$ direction, but it markedly differs from an exponential distribution in $x$  direction, thus indicating that cells do not migrate as independent units. The existence of a cell-cell cooperation is also indicated by the analysis of angle correlations between cell pairs, $\hat{\theta}(R)$, which decays with the intercellular distance $R$, see Fig.~\ref{figA}E, and by the pairwise correlation coefficient between the directions of motion, $\tilde{\theta}$, whose mean is clearly different from zero, see Fig. \ref{fig:Scatter_exp2}J.  

\subsubsection{Statistical-inference analysis: dendritic-cell experiment} \label{meb_dendritic}

\begin{figure}
\begin{center}
\includegraphics[scale=1.7]{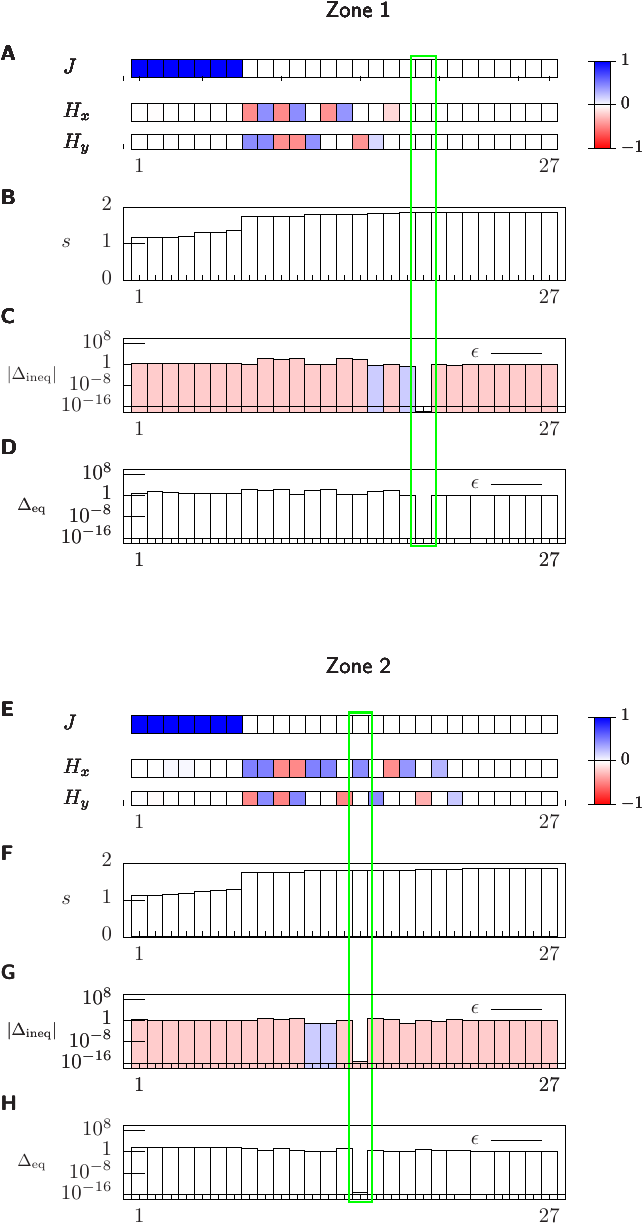}
\caption{
\textbf{Statistical-inference analysis of the dendritic-cell experiment with high cell density with the maximum-entropy method with bound constraints}. (\textbf{A})-(\textbf{D}) analysis for data in zone 1, where we use the same notation as in Fig. \ref{fig4}. (\textbf{E})-(\textbf{H})  analysis for zone 2. The numerical values of $J$ and $\bH$ for the MEb solutions are shown in Table \ref{Table2}.}
\label{fig5}
\end{center}
\end{figure}

\begin{figure}
\begin{center}
\includegraphics[scale=1.7]{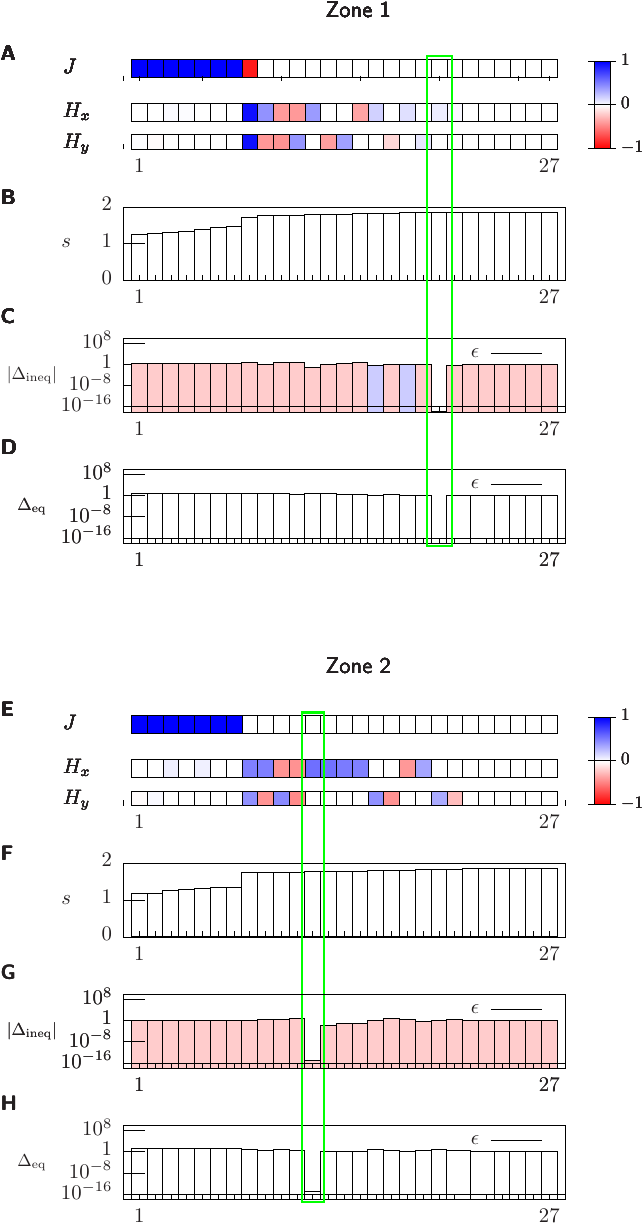}
\caption{
\textbf{Statistical-inference analysis of the dendritic-cell experiment with low cell density with the maximum-entropy method with bound constraints}. (\textbf{A})-(\textbf{D}) analysis for data in zone 1, where we use the same notation as in Fig. \ref{fig4}. (\textbf{E})-(\textbf{H})  analysis for zone 2. The numerical values of $J$ and $\bH$ for the MEb solutions are shown in Table \ref{Table2}.}
\label{fig7}
\end{center}
\end{figure}

Proceeding along the lines of the wound-healing experiment, we estimated the uncertainty on cell positions resulting from the finite pixel size as described in Section \ref{sec_delta}. This uncertainty results in broad confidence intervals for the feature averages: indeed, Table \ref{Table2} shows that the relative spread for the empirical average of the correlation $(\langle C\rangle_{\rm ex}^{\rm max}-\langle C\rangle_{\rm ex}^{\rm min})/(\langle C\rangle_{\rm ex}^{\rm max}+\langle C\rangle_{\rm ex}^{\rm min})$ can be as large as $~100 \%$, and similarly for the polarization averages, thus confirming the need for the MEb approach. 
 
Figures \ref{fig5} and \ref{fig7} show the MEb solution for the dendritic-cell experiment in the high- and low-density case and chemokine-poor and rich regions, i.e., zones 1 and 2, respectively, and the resulting values of $J$ and $\bH$ are shown in Table \ref{Table2}. 
First, we observe that the $y$ component of the external field $\bH$ is very small, and that the $x$ component vanishes in zone 1, while it is different from zero in zone 2 for all densities: this dependence of the inferred external field on the zone reflects  the chemokine gradient built in the device in the horizontal direction, see Fig. \ref{figB}.

Second, for all densities and zones, the MEb indicates that the interaction parameter, $J$, is null. 
This result is supported by the motional-data analysis of Section \ref{sec:C3}. In fact, Fig.~\ref{figB}G demonstrates  that the angle correlations between cell pairs, $\hat{\theta}(R)$, does not depend on the intercellular distance, and Figs. \ref{fig:correlations-Zone1}F and \ref{fig:Scatter3}F indicate that the pairwise correlation coefficient between the directions of motion, $\tilde{\theta}$, has zero mean. Overall, these results are markedly different from the ones obtained for the wound-healing experiment in Section \ref{meb_cancer}, and they indicate that these dendritic cells migrate individually in the chemokine gradient.

\section{Discussion}\label{Sec:Discussion}

Motivated by the recent, ubiquitous applications of inference methods to biological systems \cite{bialek2012statistical,schneidman2006weak,shlens2006the,seno2008maximum,weigt2009identification,lezon2006using} and by the 
importance of cell-migration phenomena in both physiological and pathological contexts \cite{carmona2008contact,mayor2010keeping},  we proposed a statistical-inference method to detect and single out cell-cell interactions in a population of migrating cells. 

While there is a variety of cases of collectively moving biological entities which do not have direct physical interactions, it is often unclear whether migrating cells, such as immune cells, could interact remotely, e.g., by means of diffusible factors or other intercellular signals, and thus display collective behaviors. This issue is particularly important when cells migrate in the same direction towards an external attractor, e.g., a biochemical signal: in fact, it is difficult to tell whether cells all go in the same direction independently of each other simply because they are all attracted by the same cue, or if they also interact with each other through some signals. 

To adress this question, we proposed a statistical-inference method specifically designed for cell-tracking experiments, which is capable of  handling the empirical uncertainties specific to tracking processes, such as the errors resulting from finite camera resolution, missing tracks, and others. 
In addition, the mean-field structure of our inferred statistical model allows for an explicit  solution even for a finite number of cells---a feature which may prove to be particularly useful for, say, experiments on lab-on-a-chip technology, where the  number of tracked cells can be small. 

To check the soundness of our inference approach, we first tested it on synthetic data sets, i.e.,  trajectories generated from an  XY model of spins, and on tracks generated from a non-mean-field self-propelled model of particles.  We find, overall, a very good agreement between the original and the inferred values of the model parameters, e.g., the spin-spin couplings and the strength of the external field to which the self-propelled particles are subject.

Building on the results above,  we applied our  method to two prototypical cell-migration experiments: Mesenchymal migration towards a wound, and amoeboid migration of immune cells, i.e., dendritic cells, following a chemokine gradient. 

The inference analysis gives strong  evidence of intercellular interactions for the wound-healing experiment, which  served as a stereotypical case of collective migration, thus providing a positive control for detection of cell-cell interactions by our method.

As far as the dendritic-cell experiment is concerned, immune cells release molecules towards the extracellular milieu, which could steer the migration of adjacent cells \cite{Majumdar2019}. 
Indeed, the release of vesicles or small molecules to extracellular milieu as a mechanism of paracrine cell communication allows coordinated migration in a contact-independent manner in other cellular systems \cite {Ma2014, Kriebel2018}. Similarly, dendritic cells release ATP, which acts in an autocrine manner \cite {Pablo2017}, but it is unknown whether it can affect the migration of adjacent cells. Nonetheless, our inference method did not provide evidence of cell-cell interactions during dendritic cells chemotaxis,  revealing that these cells move independently towards the gradient. 

In this regard, it is important to point out that the statistical-inference method that we proposed detects cell-cell interactions which are instantaneous, i.e., whose propagation time is significantly shorter than all other time scales: in fact, the observables that have been chosen in the ME method, Eq. (\ref{eq_c}), involve directions of motion $\bs_i(t) \cdot \bs_j(t)$ of cells $i$ and $j$ evaluated at the same instant of time $t$, see Section \ref{mei}. Despite the fact that such instantaneous interactions have been used  in a variety of statistical-inference studies for biological systems \cite{schneidman2006weak,bialek2012statistical}, other types of interactions may be present in cell-migration experiments. For example, cell $i$ may release locally a  chemical compound along its migratory path, and cell may $j$ cross the former  path of $i$ at  a later time, and thus feel a delayed interaction with $i$ mediated by this compound. The implementation of a time-lagged ME model goes beyond the scope of this work, but it may be studied by introducing  in the present model  time-lagged  observables, in which the directions of motion  $\bs_i$ and $\bs_j$ appear at different instants of time \cite{tang2008a}. Importantly, the existence of a time-lagged interaction would denote a type of collective migration which is markedly different from the one adressed in this study. In fact, while instantaneous interactions characterize a genuinely collective behavior, time-lagged interactions would be rather an indirect effect due to the fact that cell $i$ alters the environment in which cell $j$ moves. As a result, the absence of instantaneous  cell-cell interactions resulting from our analysis indicates the absence of a genuinely collective behavior in the dendritic-cell experiment.

In order to trigger the immune response, dendritic cells need to uptake an antigen in the peripheral tissues, and then quickly reach the lymph nodes via the lymphatic vessels \cite{Saez2018}: in this regard, our statistical-inference analysis suggests that dendritic cells follow individually signals from the lymphatic vessels (i.e. CCL21 chemokine), disregarding any other signal. 
It is conceivable that such  absence of cell-to-cell interactions during dendritic-cell migration may correspond to a strategy to undergo their main function. In fact, dendritic-cell migration markedly differs from the one of other immune cells, e.g.,  neutrophils, which, when receiving an attractive signal from a pathogen, emit secondary signals to attract other neutrophils on site \cite{futosi2013neutrophil}. For dendritic cells, only the cells activated directly  by the pathogen should leave the tissue to reach the lymph nodes: as a result, the fact that these cells should not emit such  signal to attract other cells allows for a natural interpretation of our statistical-inference results.
Finally, the individual migratory mechanism that we inferred may allow the cells to disregard not only intercellular, but also external signals that might delay their arrival  towards the lymphatic vessels and ultimately to the lymph nodes, thus constituting a strategy to reach the vessels as efficiently as possible. Further studies should be done in this direction to corroborate these hypotheses.

\bibliographystyle{plain}

\section*{Acknowledgments}
E.A., A.B., M.C., M.P. and P.V. conceived the study. E.A. and M.C. developed the theory. P.J.S. designed and performed the experiments. E.A., A.B. and M.C. wrote most of the paper, P.J.S. and M.P. contributed to the revisions. We thank L. Caccianini, G. Colazzo, L. Del Mercato,  J.-F. Joanny, A. S. Kumar, G. Maruccio,  P. \v{S}ulc and A. Zilman for valuable conversations.

\clearpage


\section*{Supplementary material}

# Acknowledgments

# Supplementary Material for "A statistical-inference approach to reconstruct inter-cellular interactions in cell-migration experiments"

Elena Agliari[1], Pablo J. Sáez[2], Adriano Barra[3], Matthieu Piel[2], Pablo Vargas[2], and Michele Castellana [*4,5]

[1]Dipartimento di Matematica, Sapienza Università di Roma, Rome, Italy
[2]UMR 144, Institut Curie, Paris, France
[3]Dipartimento di Matematica & Fisica 'Ennio De Giorgi', Università del Salento, Lecce, Italy
[4]Laboratoire Physico-Chimie Curie, Institut Curie, PSL Research University, CNRS UMR 168, Paris, France
[5]Sorbonne Universités, UPMC Univ. Paris 06, Paris, France

## Contents

# S1 Maximum-entropy models

## S1.1 Maximum-entropy methods with equality constraints

In this Section, we will review maximum-entropy models with equality constraints [6], which rely on the hypothesis that the experimental data provide exact estimates for the correlation and polarization. In particular, here we assume that the nominal cell positions at different times are all different, i.e., all directions of motion $\mathbf{s}_i(t)$ are well defined. For details on the derivation refer to Section S1.2, where we present the extension to inequality constraints.

We denote the exact experimental estimate for the correlation and polarization by

$$\langle C(\mathbf{S}) \rangle = C_{\text{ex}}, \tag{S1}$$

$$\langle \boldsymbol{M}(\mathbf{S}) \rangle = \boldsymbol{M}_{\text{ex}}, \tag{S2}$$

where $\langle \rangle$ is the average over a set of temporal snapshots of the population, and we estimate the minimal probability distribution of the velocities with the ME principle given by Eqs. (4), (5) (6):

$$\max_P S[P] \tag{S3}$$

subject to

$$\langle C(\mathbf{S}) \rangle_P = C_{\text{ex}}, \tag{S4}$$

$$\langle \boldsymbol{M}(\mathbf{S}) \rangle_P = \boldsymbol{M}_{\text{ex}}, \tag{S5}$$

$$\int d\mathbf{S} P(\mathbf{S}) = 1, \tag{S6}$$

where the entropy $S$ is given by the Eq. (3) and the observable $x$ by the directions of motion $\mathbf{S}$. In the equations above, $\int d\mathbf{S} \equiv \int d\mathbf{s}_1 \cdots d\mathbf{s}_N$ denotes the integral over the directions

$$\mathbf{s}_i = (\cos\theta_i, \sin\theta_i), \tag{S7}$$

where $\int d\mathbf{s}_i \equiv \int_0^{2\pi} d\theta_i$, and $\langle \cdot \rangle_P \equiv \int d\mathbf{S} P(\mathbf{S}) \cdot$ is the average with the model distribution $P$. The solution of the optimization problem (S3)-(S6) is given by Eq. (11), where the partition function reads

$$Z = \int d\mathbf{S} \exp\left\{N[JC(\mathbf{S}) + \boldsymbol{H} \cdot \boldsymbol{M}(\mathbf{S})]\right\}. \tag{S8}$$

The average correlation and polarization can be obtained explicitly for any given $J$ and $\boldsymbol{H}$. Setting

$$\kappa_{\pm} \equiv \sqrt{\frac{\pm 2J}{N-1}}, \tag{S9}$$

for $J > 0$, the result is

$$\langle C(\mathbf{S}) \rangle_P = -\frac{1}{N-1} + \frac{(N-1)|\boldsymbol{H}|^2}{4NJ^2} \tag{S10}$$

$$+ \frac{\int_0^\infty r\,dr\, e^{-NS_+(r)}\left[\frac{\mathrm{I}_1(\sqrt{2J}r)r}{\sqrt{2J}\mathrm{I}_0(\sqrt{2J}r)}\mathrm{I}_0\left(\frac{\sqrt{N-1}}{\kappa_+}r\,|\boldsymbol{H}|\right) - \mathrm{I}_1\left(\frac{\sqrt{N-1}}{\kappa_+}r\,|\boldsymbol{H}|\right)\frac{(N-1)r\,|\boldsymbol{H}|}{N(2J)^{3/2}}\right]}{\int_0^\infty r\,dr\, e^{-NS_+(r)}\mathrm{I}_0\left(\frac{\sqrt{N-1}}{\kappa_+}r\,|\boldsymbol{H}|\right)},$$

$$\langle \boldsymbol{M}(\mathbf{S}) \rangle_P = -\frac{(N-1)\boldsymbol{H}}{2NJ} + \frac{\int_0^\infty r\,dr\, e^{-NS_+(r)}\frac{(N-1)r\boldsymbol{H}}{N\sqrt{2J}\,|\boldsymbol{H}|}\mathrm{I}_1\left(\frac{\sqrt{N-1}}{\kappa_+}r\,|\boldsymbol{H}|\right)}{\int_0^\infty r\,dr\, e^{-NS_+(r)}\mathrm{I}_0\left(\frac{\sqrt{N-1}}{\kappa_+}r\,|\boldsymbol{H}|\right)}, \tag{S11}$$

where

$$S_+(r) \equiv \frac{N-1}{2N}r^2 - \log[\mathrm{I}_0(\sqrt{N-1}\kappa_+ r)], \tag{S12}$$

and $I_n$ is the modified Bessel function of the first kind [1].

For $J < 0$, we obtain

$$\langle C(\mathbf{S}) \rangle_P = -\frac{1}{N-1} + \frac{(N-1)|\boldsymbol{H}|^2}{4NJ^2} \tag{S13}$$

$$+ \frac{\int_0^\infty r dr e^{-NS_-(r)} \left[ \frac{J_1(\sqrt{-2J}r)r}{\sqrt{-2J}J_0(\sqrt{-2J}r)} J_0\left(\frac{\sqrt{N-1}}{\kappa_-}r|\boldsymbol{H}|\right) - J_1\left(\frac{\sqrt{N-1}}{\kappa_-}r|\boldsymbol{H}|\right) \frac{(N-1)r|\boldsymbol{H}|}{N(-2J)^{3/2}} \right]}{\int_0^\infty r dr e^{-NS_-(r)} J_0\left(\frac{\sqrt{N-1}}{\kappa_-}r|\boldsymbol{H}|\right)},$$

$$\langle \boldsymbol{M}(\mathbf{S}) \rangle_P = -\frac{(N-1)\boldsymbol{H}}{2NJ} - \frac{\int_0^\infty r dr e^{-NS_-(r)} \frac{(N-1)r\boldsymbol{H}}{N\sqrt{-2J}|\boldsymbol{H}|} J_1\left(\frac{\sqrt{N-1}}{\kappa_-}r|\boldsymbol{H}|\right)}{\int_0^\infty r dr e^{-NS_-(r)} J_0\left(\frac{\sqrt{N-1}}{\kappa_-}r|\boldsymbol{H}|\right)}, \tag{S14}$$

where

$$S_-(r) \equiv \frac{N-1}{2N}r^2 - \log[J_0(\sqrt{N-1}\kappa_- r)], \tag{S15}$$

and $J_n$ is the Bessel function of the first kind.

Finally, for $J = 0$ the velocities behave as independent variables, and the correlation factors out as a product of polarizations:

$$\langle C(\mathbf{S}) \rangle_P = |\langle \boldsymbol{M}(\mathbf{S}) \rangle_P|^2, \tag{S16}$$

$$\langle \boldsymbol{M}(\mathbf{S}) \rangle_P = \frac{I_1(|\boldsymbol{H}|)\boldsymbol{H}}{I_0(|\boldsymbol{H}|)|\boldsymbol{H}|}. \tag{S17}$$

## S1.2   Maximum-entropy methods with bound constraints

In this Section we discuss maximum-entropy models with bound constraints [8, 4], which apply to a more general scenario where the experimental data do not provide exact values for the correlation and polarization, but their lower and upper bounds:

$$C_{\text{ex}}^{\min} \leq \langle C(\mathbf{S}) \rangle \leq C_{\text{ex}}^{\max} \tag{S18}$$

$$\boldsymbol{M}_{\text{ex}}^{\min} \leq \langle \boldsymbol{M}(\mathbf{S}) \rangle \leq \boldsymbol{M}_{\text{ex}}^{\max}. \tag{S19}$$

We construct the minimal probability distribution of the velocities with the maximum-entropy principle:

$$\max_P S[P] \tag{S20}$$

subject to

$$C_{\text{ex}}^{\min} \leq \langle C(\mathbf{S}) \rangle_P \leq C_{\text{ex}}^{\max}, \tag{S21}$$

$$\boldsymbol{M}_{\text{ex}}^{\min} \leq \langle \boldsymbol{M}(\mathbf{S}) \rangle_P \leq \boldsymbol{M}_{\text{ex}}^{\max}, \tag{S22}$$

$$\int d\mathbf{S} P(\mathbf{S}) = 1. \tag{S23}$$

The optimization problem (S20)-(S23) involves both equality and inequality constraints, and can be solved by means of the KKT conditions [7, 9]. To achieve this, we introduce the auxiliary function

$$L \equiv S[P] - N \left[ J_-(C_{\text{ex}}^{\min} - \langle C(\mathbf{S}) \rangle_P) + J_+(\langle C(\mathbf{S}) \rangle_P - C_{\text{ex}}^{\max}) \right.$$
$$\left. + \boldsymbol{H}_- \cdot (\boldsymbol{M}_{\text{ex}}^{\min} - \langle \boldsymbol{M}(\mathbf{S}) \rangle_P) + \boldsymbol{H}_+ \cdot (\langle \boldsymbol{M}(\mathbf{S}) \rangle_P - \boldsymbol{M}_{\text{ex}}^{\max}) + \nu \left( \int d\mathbf{S} P(\mathbf{S}) - 1 \right) \right], \tag{S24}$$

where $J_\pm$, $H^\pm$ and $\nu$ are the Lagrange multipliers related to the inequality constraints (S21) and (S22) and to the equality constraint (S23), respectively. Note that we defined the multipliers by including the additional factor $N$ in such a way that these are of order unity for large $N$. The KKT conditions are given by the stationarity condition for the auxiliary function

$$0 = \frac{\partial L}{\partial P(\mathbf{S})} = -\log P(\mathbf{S}) - 1 - N[JC(\mathbf{S}) + \boldsymbol{H} \cdot \boldsymbol{M}(\mathbf{S}) + \nu], \tag{S25}$$

where we have set

$$J \equiv J_+ - J_-, \tag{S26}$$
$$\boldsymbol{H} \equiv \boldsymbol{H}_+ - \boldsymbol{H}_-. \tag{S27}$$

In addition, the optimum must satisfy the primal-feasibility conditions (S21), (S22) and (S23), the dual-feasibility conditions

$$J_\pm \geq 0,$$
$$\boldsymbol{H}_\alpha^\pm \geq 0, \ \alpha = x, y \tag{S28}$$

and the complementary-slackness conditions

$$\pm J_\pm(\langle C(\mathbf{S})\rangle_P - C_{\mathrm{ex}}^{\mathrm{max(min)}}) = 0,$$
$$\pm \boldsymbol{H}_\alpha^\pm(\langle \boldsymbol{M}_\alpha(\mathbf{S})\rangle_P - \boldsymbol{M}_\alpha^{\mathrm{max(min)}}) = 0, \tag{S29}$$

where, in what follows, the identities that contain a $\pm$ sign denote two distinct equations, one for each value of the sign, and the index $\alpha$ runs over the two vector components. By solving Eq. (S25) for $P$ and using the normalization condition (S23), we obtain the explicit form (11) for the distribution of normalized velocities, where the partition function is given by Eq. (S8).

### S1.2.1  Partition function

We will now compute the partition function, so as to obtain an explicit expression for the averages $\langle\rangle_P$ that appear in the KKT conditions. First, we rewrite $Z$ as a function of the sum of the directions of motion as

$$Z = \exp\left(-\frac{NJ}{N-1}\right) \int d\mathbf{S} \prod_\alpha \exp\left[\frac{J}{N-1}\left(\sum_i \mathbf{s}_{i\,\alpha}\right)^2 + H_\alpha \sum_i \mathbf{s}_{i\,\alpha}\right]. \tag{S30}$$

Let us consider the case $J > 0$ first, and rewrite the square in the exponential of Eq. (S30) in terms of a Gaussian integral [12]:

$$Z = \exp\left(-\frac{NJ}{N-1}\right) \int \frac{d\boldsymbol{x}}{2\pi} \exp\left[-\frac{1}{2}\left(\boldsymbol{x} - \frac{\boldsymbol{H}}{\kappa_+}\right)^2\right][f(\boldsymbol{x})]^N, \tag{S31}$$

where $\boldsymbol{x} = (x_1, x_2)$, $\kappa_\pm$ is given by Eq. (S9), we set

$$f(\boldsymbol{x}) \equiv \int_0^{2\pi} d\theta \exp[\kappa_+(x_1\cos\theta + x_2\sin\theta)] = 2\pi \mathrm{I}_0(\kappa_+|\boldsymbol{x}|), \tag{S32}$$

and $\mathrm{I}_n$ is the modified Bessel function of the first kind [1]. By using polar coordinates $\boldsymbol{x} = r(\cos\phi, \sin\phi)$ in the integral in Eq. (S31), we obtain

$$Z = \frac{N-1}{2\pi} \exp\left(-\frac{NJ}{N-1} - \frac{|\boldsymbol{H}|^2}{2\kappa_+^2}\right)$$
$$\times \int_0^\infty r\,dr\,e^{-NS_+(r)} \int_0^{2\pi} d\phi \exp\left[\frac{\sqrt{N-1}}{\kappa_+}r(\boldsymbol{H}_1\cos\phi + \boldsymbol{H}_2\sin\phi)\right]$$
$$= (2\pi)^N(N-1)\exp\left(-\frac{NJ}{N-1} - \frac{(N-1)|\boldsymbol{H}|^2}{4J}\right)\int_0^\infty r\,dr\,e^{-NS_+(r)}\mathrm{I}_0\left(\frac{\sqrt{N-1}}{\kappa_+}r|\boldsymbol{H}|\right), \tag{S33}$$

where we have defined $S_+$ according to Eq. (S12), and in the second line we rewrote the integral with respect to $\phi$ in terms of $I_0$, cf. Eq. (S32). In Eq. (S33), we have reduced the partition function to a simple, one-dimensional integral that can be evaluated with arbitrary precision. In addition, the average correlation and polarization can be obtained from (S33) by taking its derivatives with respect to the Lagrange multipliers:

$$
\langle C(\mathbf{S}) \rangle_P = \frac{1}{N} \frac{\partial \log Z}{\partial J}, \tag{S34}
$$

$$
\langle \boldsymbol{M}(\mathbf{S}) \rangle_P = \frac{1}{N} \frac{\partial \log Z}{\partial \boldsymbol{H}}, \tag{S35}
$$

and the result is given by Eqs. (S10) and (S11).

Proceeding along the same lines for $J < 0$, we obtain Eqs. (S13) and (S14), where $S_-$ is given by Eq. (S15), and $J_n$ is the Bessel function of the first kind [1]. Finally, we consider the case $J = 0$, for which the directions $\mathbf{s}_i$ in $P(\mathbf{S})$ behave as independent variables, and the average correlation and polarization are given by Eq.s (S16) and (S17).

Overall, Eqs. (S10), (S11), (S13), (S14), (S16) and (S17), yield the average correlation and polarization for any value of the KKT multipliers, thus allowing us to solve the KKT conditions. In particular, the solution above for the average correlation and polarization is exact for any value of $N$, thus allowing us to study both small and large cell populations.

### S1.2.2 Solution of Karush-Kuhn-Tucker conditions and algorithmic implementation

Hereafter we describe the strategy used to solve the KKT conditions. According to Eq. (S29), the KKT multipliers are nonnegative: It follows that for any pair of maximal and minimal bound constraints, e.g., Eq. (S21), the complementary-slackness conditions (S29) imply three possible cases:

- $J_+ > 0$, $J_- = 0$, thus Eq. (S29) implies $\langle C(\mathbf{S}) \rangle_P = C_{\mathrm{ex}}^{\mathrm{max}}$,

- $J_+ = 0$, $J_- > 0$, thus $\langle C(\mathbf{S}) \rangle_P = C_{\mathrm{ex}}^{\mathrm{min}}$,

- $J_+ = J_- = 0$,

where the case $J_+ > 0$, $J_- > 0$ would imply $\langle C(\mathbf{S}) \rangle_P = C_{\mathrm{ex}}^{\mathrm{max}} = C_{\mathrm{ex}}^{\mathrm{min}}$, and is ruled out because we assume that $C_{\mathrm{ex}}^{\mathrm{max}} > C_{\mathrm{ex}}^{\mathrm{min}}$, and similarly for the other constraints. In the three cases above, each multiplier $J_+$, $J_-$ may be either equal to zero or positive, and in the latter case the multiplier is determined by the condition $\langle C(\mathbf{S}) \rangle_P = C_{\mathrm{ex}}^{\mathrm{max}}$ or $\langle C(\mathbf{S}) \rangle_P = C_{\mathrm{ex}}^{\mathrm{min}}$. As a result, each case uniquely determines both multipliers.

In order to solve the full KKT conditions, we considered the three cases above for each of the three pairs of maximal and minimal bound constraints, cf. Eqs. (S21) and (S22), and obtained a total of twenty-seven cases. We solved the KKT conditions for the multipliers in each of these cases as discussed above, and checked whether the solution satisfies the complementary-slackness conditions (S29) and the primal-feasibility conditions (S21) and (S22).

First, the fulfillment of the equality conditions (S29) is assessed as follows. For each nonzero multiplier, there is an equality condition that needs to be satisfied. For example, if $J_+ > 0$, $J_- = 0$ and $\boldsymbol{H}^+ = \boldsymbol{H}^- = \mathbf{0}$, the only equality condition is $\langle C(\mathbf{S}) \rangle_P = C_{\mathrm{ex}}^{\mathrm{max}}$. We thus introduce the relative residual associated with this condition:

$$
\Delta_{\mathrm{eq}} = \left| \frac{\langle C(\mathbf{S}) \rangle_P - C_{\mathrm{ex}}^{\mathrm{max}}}{C_{\mathrm{ex}}^{\mathrm{max}}} \right|. \tag{S36}
$$

If there are multiple equality conditions, $\Delta_{\mathrm{eq}}$ is defined as the maximum over the residuals of all equality conditions. Second, the fulfillment of the inequality conditions (S21) and (S22) is assessed by introducing the residual

$$
\Delta_{\mathrm{ineq}} = \min \left\{ \frac{\langle C(\mathbf{S}) \rangle_P - C_{\mathrm{ex}}^{\mathrm{min}}}{|C_{\mathrm{ex}}^{\mathrm{min}}|}, \frac{C_{\mathrm{ex}}^{\mathrm{max}} - \langle C(\mathbf{S}) \rangle_P}{|C_{\mathrm{ex}}^{\mathrm{max}}|}, \cdots \right\}, \tag{S37}
$$

where the first two terms in braces are the relative residuals of Eq. (S21), and $\cdots$ incorporates the analog for Eqs. (S22). If, for a given case under consideration, $\Delta_{\text{eq}} = 0$ and $\Delta_{\text{ineq}} \geq 0$ within numerical precision, then the equality and inequality conditions (S29), (S21) and (S22) are satisfied, and we consider the case as admissible, see Fig. 5B, 6B and 4.

Finally, the solution of the ME problem is given by the admissible case with the largest entropy, see Figs. 5, 6 and 4.

## S2 Tests of the maximum-entropy method with bounds on synthetic data

To test the predictive capabilities of the MEb method, we generate synthetic data for a system of $N$ elements which evolve according to a given dynamics. We will consider two different models for the synthetic dynamics: The XY model and a model of SP particles. We denote by $\mathcal{P} = \{p_1, ..., p_k\}$ the set of parameters which determine the dynamics, generate samples of the configurations of the $N$ elements for different choices of $\mathcal{P}$, and analyzed them with the MEb.

### S2.1 XY model

The XY model is a statistical-mechanical model originally introduced to describe ferromagnetic systems [5]. The model consists of $N$ spins $\mathbf{S} = \{\mathbf{s}_1, ..., \mathbf{s}_N\}$ which interact pairwise, and which can be written in terms of their orientation angle $\theta$ by means of Eq. (S7). The Hamiltonian for this system is given by Eq. (12).

Notice that the larger $J$, and the higher the energy cost for $\mathbf{s}_i$ and $\mathbf{s}_j$ to be misaligned; similarly, the larger $\boldsymbol{H}$, the higher the energy cost for $\mathbf{s}_i$ to misalign with respect to the direction of the external field.

For a given set of parameters $\mathcal{P} = \{J, \boldsymbol{H}\}$, we initialize the system in a configuration $\mathbf{S}$ and let it evolve with a single-spin-flip Glauber dynamics, reaching equilibration after a large enough number of spin flips. Then, we collect $T = 100$ configuration samples $\mathbf{S}_1, \cdots, \mathbf{S}_T$, with which we compute the configuration averages $\langle C(\mathbf{S}) \rangle_{\text{ex}}$ and $\langle \boldsymbol{M}(\mathbf{S}) \rangle_{\text{ex}}$, cf. Eq. (1). Given that the averages above are estimated from a finite number of samples, they are subjected to statistical error. We thus estimate the confidence interval for these averages, i.e., the lower and upper bounds in Eqs. (S18) and (S19), in terms of this statistical error, which we denote by $\sigma$, setting $C_{\text{ex}}^{\text{max(min)}} = \langle C(\mathbf{S}) \rangle_{\text{ex}} \pm \sigma$. Finally, we use these bounds in the MEb method, and obtain the inferred values of the interaction and external field, $J_{\text{inf}}$ and $\boldsymbol{H}_{\text{inf}}$, respectively.

The procedure above is repeated $Q = 100$ times for each parameter configuration, resulting in an average and standard deviation for $J_{\text{inf}}$, $\boldsymbol{H}_{\text{inf}}$, see Fig. S1: overall, the agreement between the inferred and the original parameter values is very good, even for small $N$.

### S2.2 Self-propelled model

In the SP model, we consider $N$ particles moving on a plane. The dynamics of particles' velocities and positions is set by the following rule:

$$\mathbf{v}_i(t + \Delta t) = v_0 \Theta \left[ \alpha \sum_{j \in n_c^i} \mathbf{v}_j(t) + \beta \sum_{j \in n_c^i} \mathbf{f}_{ij} + \gamma \mathbf{h} + n_c \boldsymbol{\eta}_i(t) \right], \tag{S38}$$

$$\mathbf{x}_i(t + \Delta t) = \mathbf{x}_i(t) + \mathbf{v}_i(t) \Delta t, \tag{S39}$$

where $\Theta$ is a normalization operator $\Theta(\mathbf{y}) = \mathbf{y}/|\mathbf{y}|$ which keeps the velocity modulus fixed at $|\mathbf{v}| = v_0$, and the sum in Eq. (S38) runs over the $n_c$ nearest neighbors of particle $i$, according to a topological definition of distance. The first term in the square bracket in the right-hand side of Eq. (S38) makes particles that are close to each other move in the same direction, the second term mimics an interaction potential which accounts for excluded volume (*vide infra*), and the third term represents a uniform gradient in the direction of the unit vector $\mathbf{h}$. The fourth term represents noise, and includes a random unit vector $\eta_i(t)$, which is independent for each particle and instant of

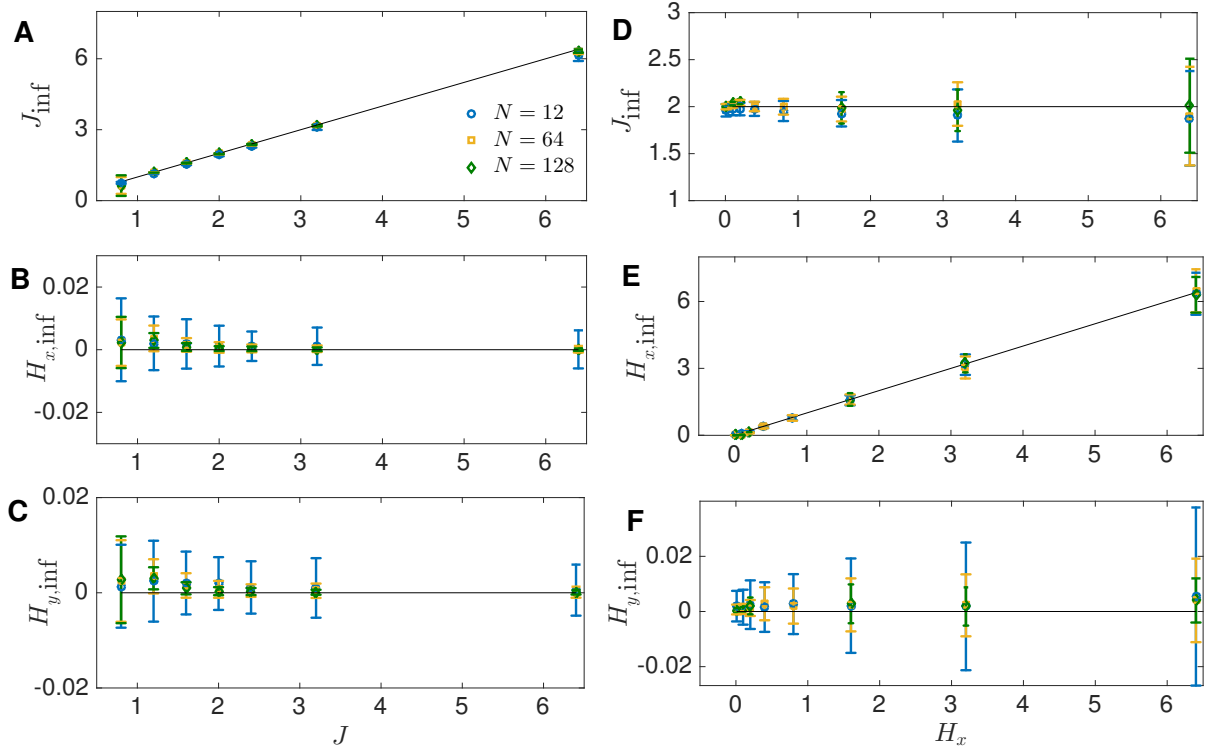

**Figure S1: Test of maximum-entropy method with bound constraints on synthetic data for the XY model.** In (**A**), (**B**) and (**C**) we set $H_x = H_y = 0$ and show the inferred parameters $J_{\mathrm{inf}}$, $\boldsymbol{H}_{\mathrm{inf}}$ as functions of $J$. In (**D**), (**E**) and (**F**) we set $J = 2$, $H_y = 0$ and show the inferred parameters as function of $H_x$. The average and standard deviation of the inferred parameters are represented by circles and error bars, respectively, and solid lines represent the original values of $J$ and $\boldsymbol{H}$ used in the simulation.

time. The parameters $\alpha, \beta, \gamma$ tune the magnitude of the terms which they multiply. The distance-dependent force $\mathbf{f}_{ij}$ acts along the direction between $i$ and $j$, $\mathbf{e}_{ij} = \mathbf{r}_{ij}/|\mathbf{r}_{ij}|$, with $\mathbf{r}_{ij} = \mathbf{r}_i - \mathbf{r}_j$, and is defined as

$$
\begin{aligned}
\mathbf{f}_{ij}(r_{ij} < r_b) &= -\infty\, \mathbf{e}_{ij}, \\
\mathbf{f}_{ij}(r_b < r_{ij} < r_a) &= \frac{1}{4}\frac{r_{ij} - r_e}{r_a - r_e}\, \mathbf{e}_{ij}, \\
\mathbf{f}_{ij}(r_a < r_{ij} < r_0) &= \mathbf{e}_{ij}, \\
\mathbf{f}_{ij}(r_{ij} > r_0) &= 0,
\end{aligned}
$$

with $r_a, r_b, r_e, r_0$ suitable length scales. More precisely, $r_b$ represents the particle size, $r_0$ sets the interaction range in such a way that particles are insensitive to each other when their distance exceeds $r_0$. $r_a$ and $r_e$ set the magnitude and direction of the force $\mathbf{f}_{ij}$, which is attractive and repulsive when $r_{ij} > r_e$ and $r_{ij} < r_e$, respectively.

We set the parameters $\alpha, \beta, \gamma, n_c, r_a, r_b, r_e, r_0$ and the direction of the field $\mathbf{h}$, and then we initialize the system by placing all the particles in an area $[x', y'] \times [x'', y'']$ and by associating to each particles a (normalized) velocity given by the angle $\theta_i, i = 1, ..., N$ randomly and uniformly drawn in $[0, 2\pi]$. Next, we let the particles move according to Eqs. (S38) and (S39), and, after a sufficient number of iterations, we collect the configurations of velocities $\mathbf{v}_i(t)$, $t = 1, \cdots, T$, and the related normalized velocities, which are then used in the MEb method.

We would like to stress that, unlike the XY model where the Hamiltonian has the same form and parameter structure as the MEb model, Eq. (12), for the SP model a direct comparison between $\mathcal{P} = \{\alpha, \beta, \gamma\}$ and the output $J_{\mathrm{inf}}$, $\boldsymbol{H}_{\mathrm{inf}}$ of the MEb is no longer possible. In particular, the former model is built out of mean-field interactions, while the latter involves a non-mean-field interaction structure. Despite the fact that there is no one-to-one mapping between the parameters of the two models, the interaction parameter $J$ in the MEb Hamiltonian (12) can be related

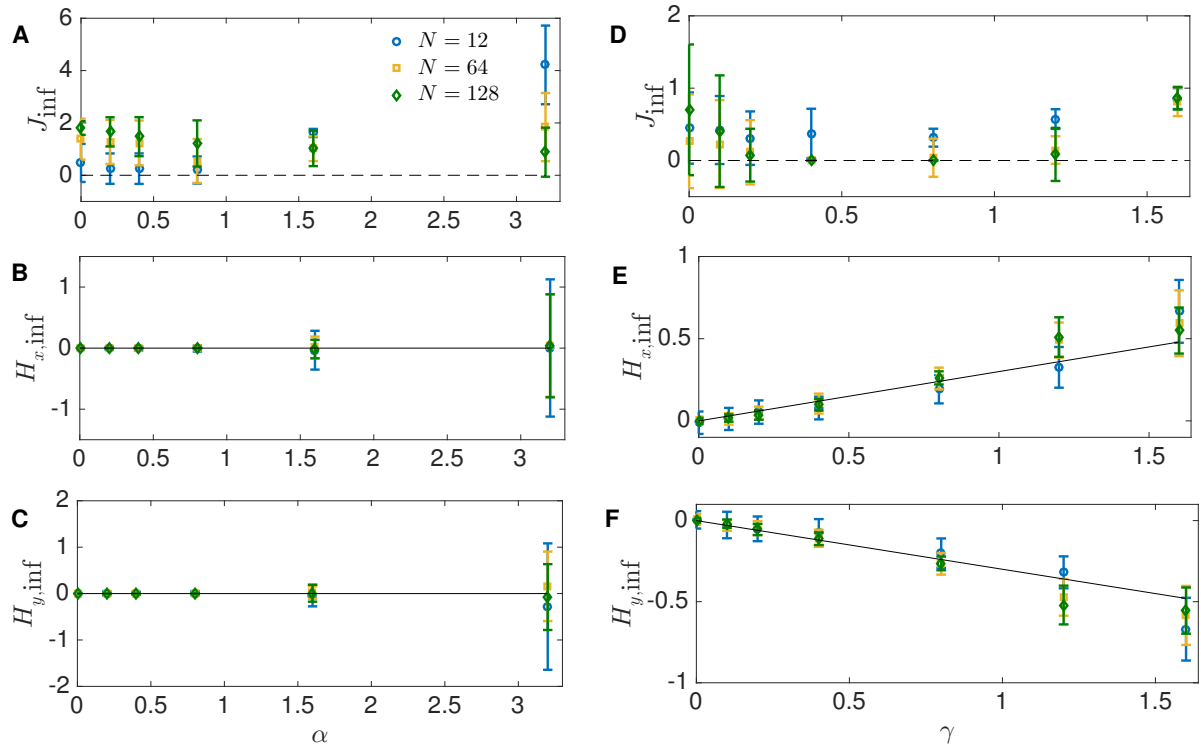

**Figure S2: Test of maximum-entropy method with bound constraints on synthetic data for the self-propelled model.** Simulations for the self-propelled model were carried on with $T = 100$ samples, and for different choices of the parameters $\alpha$ and $\gamma$, with $\beta = 0.8$, $T_0 = 50$, $n_c = 9$, $r_a = 10$, $r_b = 1$, $r_e = 1.5$ and $r_0 = 20$. In addition, we set $h_x = -h_y = 1/\sqrt{2}$, i.e., the field points in the direction $-\pi/4$. For each parameter configuration, we repeat the $Q = 100$ times. For each realization we evaluate averages, correlations and the related bounds, which are then used as input for the maximum-entropy method with bound constraints. We thus obtain $Q$ estimates of the inferred parameters $J_{\text{inf}}$, $\boldsymbol{H}_{\text{inf}}$ whose average (symbols) and standard deviations (error bars) are depicted. Results obtained with $\gamma = 0$ and varying $\alpha$ are shown in (**A**), (**B**), and (**C**), while those obtained with $\alpha = 1.2$ and varying $\gamma$ are shown in (**D**), (**E**) and (**F**). The black solid lines give the best linear fit, i.e., $y = 0$ in (**B**) and (**C**), $y = 0.3\,x$ in (**E**) and (**F**). Notice that, as the size $N$ is varied, the width of the initial area $[x', y'] \times [x'', y'']$ is modified in such a way that the initial particle density remains equal to $10^{-4}$.

to the parameter $\alpha$ in the SP dynamics, which sets the strength of the alignment between particle velocities. In fact, the inferred value of $J$ appears to be positively correlated with $\alpha$, see Fig. S2, indicating that the MEb model correctly captures the tendency of self-propelled particles to align their directions of motion. More precisely, in Fig. S2, we set $\alpha = 0$ and vary $\alpha$, and we correctly infer a vanishing field $\boldsymbol{H}$, while the inferred value of $J$ is non null. However, $J_{\text{inf}}$ seems to depend also on $N$ and its dependence on $\alpha$ appears to be non-linear. On the other hand, by setting $\alpha = 1.2$ and varying $\gamma$ by fixing $\boldsymbol{h}$ directed in the direction $-\pi/4$, we correctly infer a linear relation between $\boldsymbol{H}_{\text{inf}}$ and $\gamma$, while $J_{\text{inf}}$ still exhibits a slight dependence on $N$ and $\gamma$.

Finally, we present an additional test of the MEb method. For a given parameter configuration $\alpha = 1.2$, $\beta = 4$, $\gamma = 2.2$ and $n_c = 9$, we performed $Q = 100$ simulations of the SP with $N = 60$ particles and collected $T = 360$ snapshots of their positions. In an effort to emulate the positional uncertainty due to the camera pixels in the experiments above, for each simulation we assumed that the positions are affected by an error $\sigma = 0.02$, and rounded off the cell positions according to this error, see Section S6 for details. By using this positional uncertainty, we obtained the lower and upper bounds for the feature averages $\langle C \rangle$ and $\langle \boldsymbol{M} \rangle$, which we used to infer $J$ and $\boldsymbol{H}$ with the MEb. Furthermore, we considered the connected correlation function between motional correlation and polarization $\langle CM \rangle - \langle C \rangle \langle \boldsymbol{M} \rangle$. We compute this quantity both from inferred the MEb model above and from the data, and compare them in Fig. S3. Importantly, the MEb model is based on the average

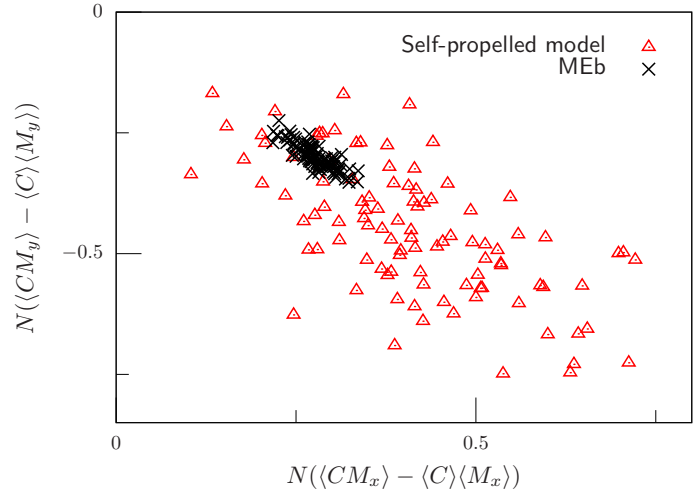

**Figure S3: Test of maximum-entropy method with bound constraints (MEb).** Given a simulation for the self-propelled model, we rounded off the particle coordinates within a pixel size $\sigma$, chosen to emulate the positional uncertainty in the experiments. The resulting $x$ and $y$ components of the connected correlation function $\langle CM \rangle - \langle C \rangle \langle M \rangle$ between the motional correlation $C$ and the polarization $M$ are shown (red), where each point corresponds to a different simulation. Because of the uncertainty resulting from the pixel size $\sigma$, the connected correlation above varies within a confidence interval, and the values shown have been obtained as the mean between the lowest and upper bound of the interval. The connected correlation above predicted by the MEb (black) is also shown, where each point corresponds to a different simulation.

correlations and polarization $\langle C \rangle$ and $\langle M \rangle$ only, not on their cross correlation $\langle CM \rangle - \langle C \rangle \langle M \rangle$: as a result, the MEb prediction for the connected correlation involves no free parameters. First, Fig. S3 shows a good agreement between model and empirical connected correlation. Second, the the MEb method reproduces quantitatively the negative correlation between the $x$ and $y$ component of $\langle CM \rangle - \langle C \rangle \langle M \rangle$ across multiple simulations—the larger the $x$ component, the smaller the $y$ component. Overall, the analysis above indicates that the MEb method—which is based on averages of empirical features only—is able to describe and capture also more complex statistical properties, e.g., the cross-correlation between two of such experimental features [2].

## S3    Wound-healing experiment

For the wound-healing experiment, human cancerous epithelial cells (HeLa) from adenocarcinoma were used. HeLa cells stably transfected with H2B-mCherry that allows live imaging of the nucleus were cultured in DMEM (Dulbecco's modified eagle medium containing glutamax) 10% fetal bovine serum (FBS) and 1% penicillin streptomycin. Once confluence was reached Hela $H2B$ cells were detached with trypsin and $0.5 \times 106$ cells were plated in a Fluorodish $24\,\mathrm{h}$ prior to the experiment. Then, a monolayer was formed and 4 scratches were done (2 in the vertical axis and 2 in the horizontal one) using a $200\,\mu\mathrm{L}$ pipette tip. Medium was changed twice after the scratching to remove detached and dead cells. The dishes were left in the incubator at $37\,^\circ\mathrm{C}$ for $4\,\mathrm{hr}$ before starting the experiment. Phase contrast images were acquired every 5 min overnight with a DMi8 inverted microscope (Leica), at $37\,^\circ\mathrm{C}$ with $5\%$ $CO_2$ atmosphere using a 10X dry objective ($NA0.40$phase), binning 2. In addition to phase contrast, mCherry fluorescence was also imaged.

## S4    Dendritic-cell experiment

Bone marrow-derived dendritic cells (BMDCs) were prepared as previously described [11]. Briefly, mouse bone marrow precursors were obtained from wild type $C57/B6$ mice and were differentiated in vitro for 10 days with

granulocyte-macrophage colony stimulating factor-containing culture medium, obtained from transfected $J558$ cells. At day 10, cells were stimulated with a pulse bacterial lipopolysaccharide (LPS) $100 \, \text{ng/ml}$, as previously described [3]. The migration experiments in collagen gels were performed as previously described [10]. LPS-stimulated BMDCs were mixed with bovine collagen type I at $3 \text{mg/ml}$ and loaded in a custom-made chamber of polydimethylsiloxane. After 30 min of incubation at $37\,^{\circ}\text{C}$, gel polymerization was reached and samples were bathed with a medium containing $200 \, \text{ng/ml}$ of $CCL21$. Finally, imaging was done as for the wound-healing experiment, but the period of acquisition was 2 min.

# S5    Analysis of motional data

## S5.1    Wound-Healing experiment

Here we focus on the wound-healing experiment, for which the number of cells is $N = 1288$.

First, we notice that the distribution of the related tracks in the observation window is not uniform: Most tracks are found in the right side of the observation area, see Fig. 2A, and cells tend to move along the horizontal direction, see Fig. 2C. This is confirmed by the scatter plots in Fig. S4: panel (A) contains the scatter plot of $v_{x\,i}(t)$ versus $v_{y\,i}(t)$, for all $t = 1, ..., T_i$ and $i = 1, ..., N$; panel (B) contains the scatter plot of $v_i(t) = |\mathbf{v}_i(t)|$ versus $\theta_i(t)$ for all $i = 1, ..., N$ and $t = 1, ..., T_i$. From the former we see that positive velocities along the $x$ direction are more likely to occur than negative ones, and the latter indicates that the density of data points is higher at $\theta \sim 0$.

Further, we investigate the presence of correlations among velocities and angles. To achieve this, we consider the Pearson coefficient , which, for two time sequences $f(i, t)$ and $f(j, t)$ is defined as

$$\tilde{f}_{ij} = \frac{\sum_t [f(i,t) - \langle f_i \rangle][f(j,t) - \langle f_j \rangle]}{\sqrt{\sum_t [f(i,t) - \langle f_i \rangle]^2} \sqrt{\sum_t [f(j,t) - \langle f_j \rangle]^2}}, \tag{S40}$$

where $\langle f_i \rangle = \sum_t f(i,t)/T'$ and the sums run until $t = T' = \min[T(i), T(j)]$. Further, in order to assess time auto-correlations, we considered the auto-covariance correlation function, which, for the observable $f(i, t)$ is defined as

$$\hat{f}_i(\tau) = \frac{\sum_t [f(i, t + \tau) - \langle f_i \rangle][f(i, t) - \langle f_i \rangle]}{\sum_t [f(i, t) - \langle f_i \rangle]^2}, \tag{S41}$$

where the sums run unti $t = T(i) - \tau$. As shown in Fig. S4C-F, pairwise correlations over the whole set of cell couples display a positive skewness for all the observables considered and the related mean $(\bar{\cdot})$, median $(\text{mdn}(\cdot))$ and standard deviation $(\text{std}(\cdot))$ turn out to be $\overline{\tilde{v}_x} = 0.183$ , $\text{mdn}(\tilde{v}_x) = 0.103$, $\text{std}(\tilde{v}_x) = 0.225$; $\overline{\tilde{v}_y} = 0.088$, $\text{mdn}(\tilde{v}_y) = 0.034$, $\text{std}(\tilde{v}_y) = 0.160$; $\overline{\tilde{v}} = 0.139$, $\text{mdn}(\tilde{v}) = 0.116$, $\text{std}(\tilde{v}) = 0.202$; $\overline{\tilde{\theta}} = 0.109$ , $\text{mdn}(\tilde{\theta}) = 0.026$ , $\text{std}(\tilde{\theta}) = 0.144$. Also, the auto-covariance for $v_x$ and $v$ tend to be positive at short times, while the auto-covariance for $v_y$ and $\theta$ display basically no dependence on time and fluctuate around zero, see Fig. S4G-J.

Next, we combine the data from all tracks and times, to study the probability distributions of $v_x$, $v_y$ and $v$. The related mean, median and standard deviation are $\overline{v_x} = 0.063$ , $\text{mdn}(v_x) = 0.060$, $\text{std}(v_x) = 0.112$; $\overline{v_y} = -0.004$, $\text{mdn}(v_y) = -0.003$, $\text{std}(v_y) = 0.075$; $\overline{v} = 0.123$, $\text{mdn}(v) = 0.107$, $\text{std}(v) = 0.080$. The best fits for their histograms are shown in Fig. S5: The best fit for the distribution of $v_x$ is given by a normal distribution $\mathcal{N}(x; \mu, \sigma) = e^{-(x-\mu)^2/(2\sigma^2)}/\sqrt{2\pi\sigma^2}$ with best-fit coefficient $\mu = 0.068 \pm 0.005$ and $\sigma = 0.12 \pm 0.01$ (panel A), while each branch of the histogram of $v_y$ it is given by an exponential distribution $\text{Exp}(-\lambda)$ with best-fit coefficient $\lambda = 19 \pm 1$ (panel B). The distribution of the modulus of the velocity is log-normal $\text{Log}\mathcal{N}(x; \mu, \sigma) = e^{-(\ln x - \mu)^2/(2\sigma^2)}/\sqrt{2\pi\sigma^2 x^2}$ with best-fit coefficients $\mu = -2.22 \pm 0.02$ and $\sigma = 0.65 \pm 0.01$ (panel C). It is worth stressing the qualitative different distributions of $v_x$ and $v_y$, and the relatively fat tail of the distribution of $v$.

## S5.2    Dendritic-cell experiment

In what follows we focus on the dendritic-cell experiment, for which we perform the same analyses discussed for the wound-healing experiment.

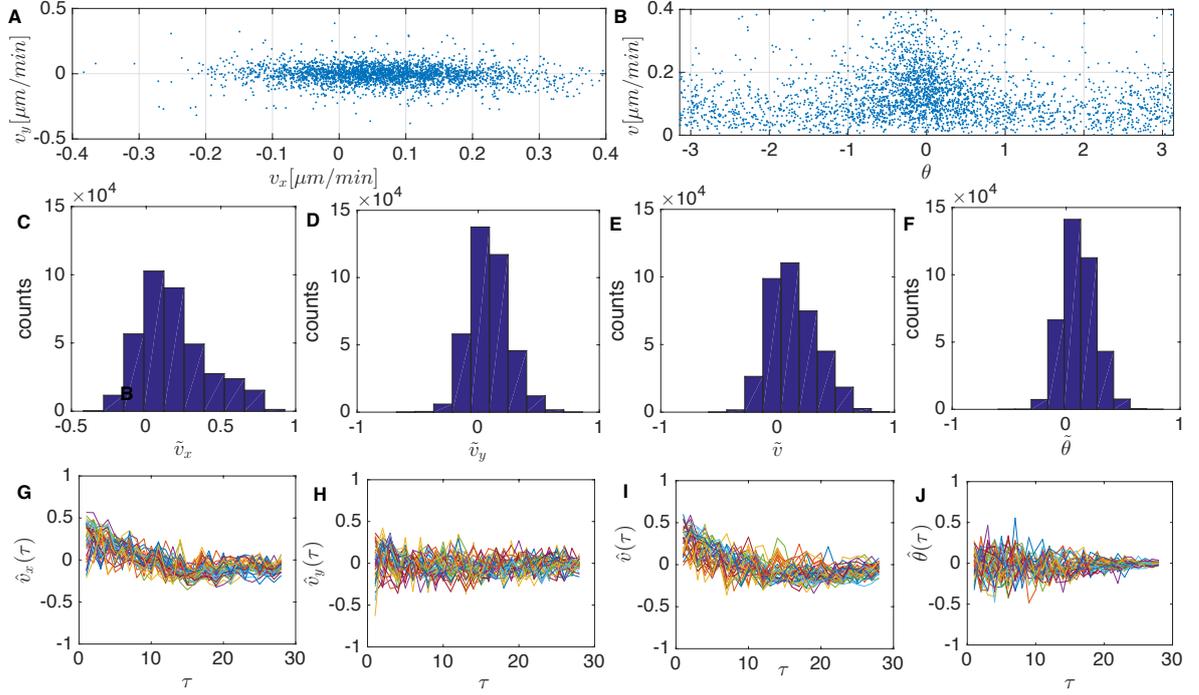

**Figure S4: Statistics of motional data for the wound-healing experiment.** First row: scatter plot for the velocity components $v_y$ vs. $v_x$ (**A**) and for the velocity norm $v$ vs. the directional angle $\theta$ (**B**). For the sake of clarity, in both panels only a 10% fraction of all available data points is shown. Second row: histograms for the pairwise correlation coefficient between the $x$-component of the velocities (**C**), between the $y$-component of the velocities (**D**), between the velocity magnitude (**E**), and between the angles (**F**). Third row: autocorrelation function for the $x$-component of the velocity (**G**), the $y$-component of the velocity (**H**), the magnitude of the velocity (**I**), and the angle (**J**). The outcomes of (a 10% fraction of) all cells are shown in different colors to give a visual representation of the variability of the autocorrelation.

### S5.2.1 Free environment (zone 1)

We consider tracks in zone 1, where the cytokine concentration is negligible. As a consequence, the cellular motion is expected to be unbiased (see Fig. 3). This kind of analysis allows us to get a picture of the migratory abilities of dendritic cells in free motion, i.e., in the absence of an external stimulus.

The overall number of tracks limited to zone 1 is $N_{\text{Low}}^{(1)} = 54$ in the low-density regime and $N_{\text{High}}^{(1)} = 231$ in the high-density regime.

First, in Fig. S6 we assess the existence of correlations between the velocity components: panel (**A**) contains the scatter plot of $v_{x\,i}(t)$ versus $v_{y\,i}(t)$, for all $i = 1, ..., N_{\text{Low}}^{(1)}$ and $t = 1, ..., T_i$; panel (**B**) contains the scatter plot of $\mathbf{v}_i(t) = |\mathbf{v}_i(t)|$ versus $\theta_i(t)$ for all $i = 1, ..., N_{\text{Low}}^{(1)}$ and $t = 1, ..., T_i$. In both cases, the data display a uniform distribution, and analogous results are obtained for the high-density case.

As far as pairwise correlations are concerned, Fig. S6C-F shows that $\tilde{v}_x, \tilde{v}_y, \tilde{v}$ and $\tilde{\theta}$ are all pretty symmetrically distributed around zero. The related mean, median and variance are $\overline{\tilde{v}_x} = 0.003$, $\text{mdn}(\tilde{v}_x) = 0.011$, $\text{std}(\tilde{v}_x) = 0.202$; $\overline{\tilde{v}_y} = 0.001$, $\text{mdn}(\tilde{v}_y) = 0.009$, $\text{std}(\tilde{v}_y) = 0.202$; $\overline{\tilde{v}} = 0.151$, $\text{mdn}(\tilde{v}) = 0.121$, $\text{std}(\tilde{v}) = 0.255$; $\overline{\tilde{\theta}} = 0.008$, $\mu(\tilde{\theta}) = 0.008$, $\text{std}(\tilde{\theta}) = 0.207$. Also, the auto-covariance for all these observables (panels G-J) display basically no dependence on time and fluctuate around zero.

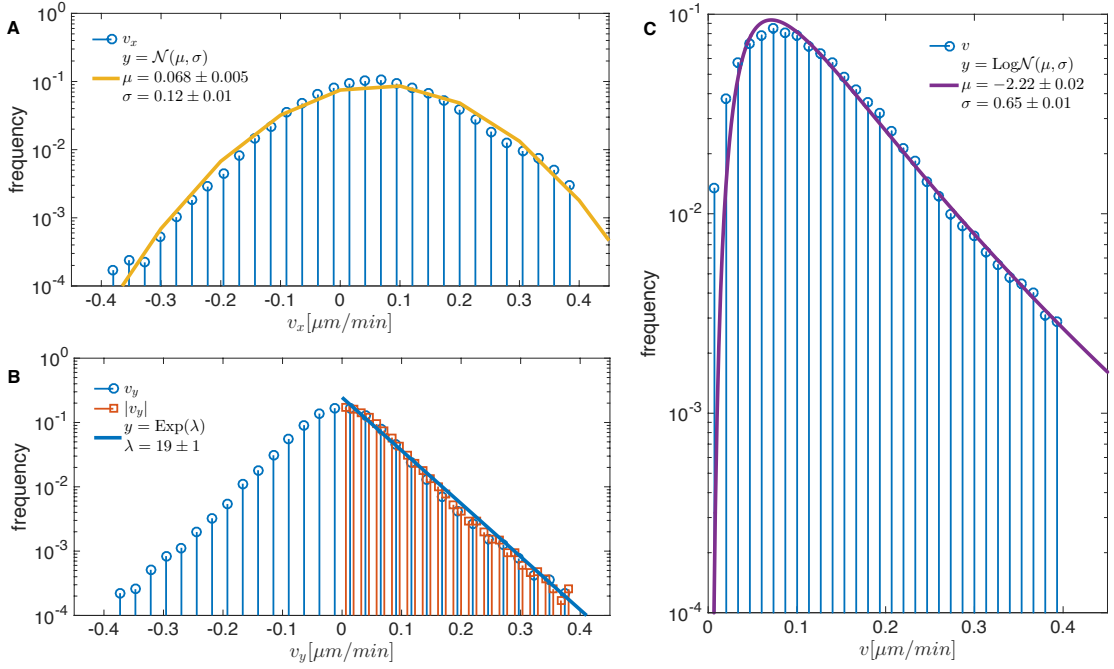

**Figure S5: Empirical distributions for cell velocities in the wound-healing experiment.** Density histogram for the $x$- and $y$-component of the velocity, panels (**A**) and (**B**), respectively, and for the magnitude $v$ (**C**) in a semi-logarithmic scale plot, with the related fit parameters. To highlight the symmetry of the distributions of $v_y$, in (**B**) we overlap the histogram for its absolute value $|v_y|$ to the positive branch of the original histogram; in fact, $v_y$ follows an exponential distribution. Finally, $v_x$ is best fitted by a normal centered at a positive value, while the velocity magnitude follows a log-normal distribution.

These results are robust with respect to the density of dendritic cells—the same conclusions are drawn in both the high- and low-density regimes—and suggest that we can look at tracks as Markov chains where at each time step we extract randomly the two components of $\mathbf{v}$ (and similarly, in a polar system, for $v$ and $\theta$). The indices of central position and dispersion are $\overline{v_x} = 0.030$ , $\mathrm{mdn}(v_x) = 0.045$, $\mathrm{std}(v_x) = 1.807$; $\overline{v_y} = 0.046$, $\mathrm{mdn}(v_y) = 0.055$, $\mathrm{std}(v_y) = 1.863$; $\overline{v} = 2.132$, $\mathrm{mdn}(v) = 1.848$, $\mathrm{std}(v) = 1.480$. The probability distributions of velocities are shown in Fig. S7: the best fit for $|v_x|$ (panel A) and $|v_y|$ (panel B) is given by an exponential distribution $\mathrm{Exp}(x; \lambda) = \lambda e^{-\lambda x}$ for which a least-square fit yields compatible values of lambda $\lambda = 0.80 \pm 0.04$ for both components. The analogy of the behavior exhibited by the two components corroborates that the motion is fully isotropic, as evidenced by the polar histogram for $\theta$, shown in Fig. 3C. As for the magnitude of the velocity (panel C), the best fitting function is provided by a Gumbel distribution $\mathrm{Gumbel}(x; \alpha, \beta) = \frac{1}{\beta} e^{-z + e^{-z}}$, where $z = (x - \alpha)/\beta$, with best fitting parameters $\alpha = 1.20 \pm 0.04$ and $\beta = 1.4 \pm 0.1$. Overall, the distributions shown in Fig. S7 are short-tailed, and thus expected to preserve the central-limit-theorem regularity in the large-sample limit.

### S5.2.2 Perturbed environment (zone 2)

We consider tracks in zone 2, where the cytokine concentration is high. The overall number of tracks limited to zone 2 is $N_{\mathrm{Low}}^{(3)} = 75$ in the low-density regime and $N_{\mathrm{High}}^{(3)} = 157$ in the high-density regime. As a consequence of the cytokine gradient, the cellular motion is expected to be biased and, in fact, this appears clear even by eye inspection (see Fig. 3). In particular, there is a bias along the $x$ direction, see the scatter plot for the velocity

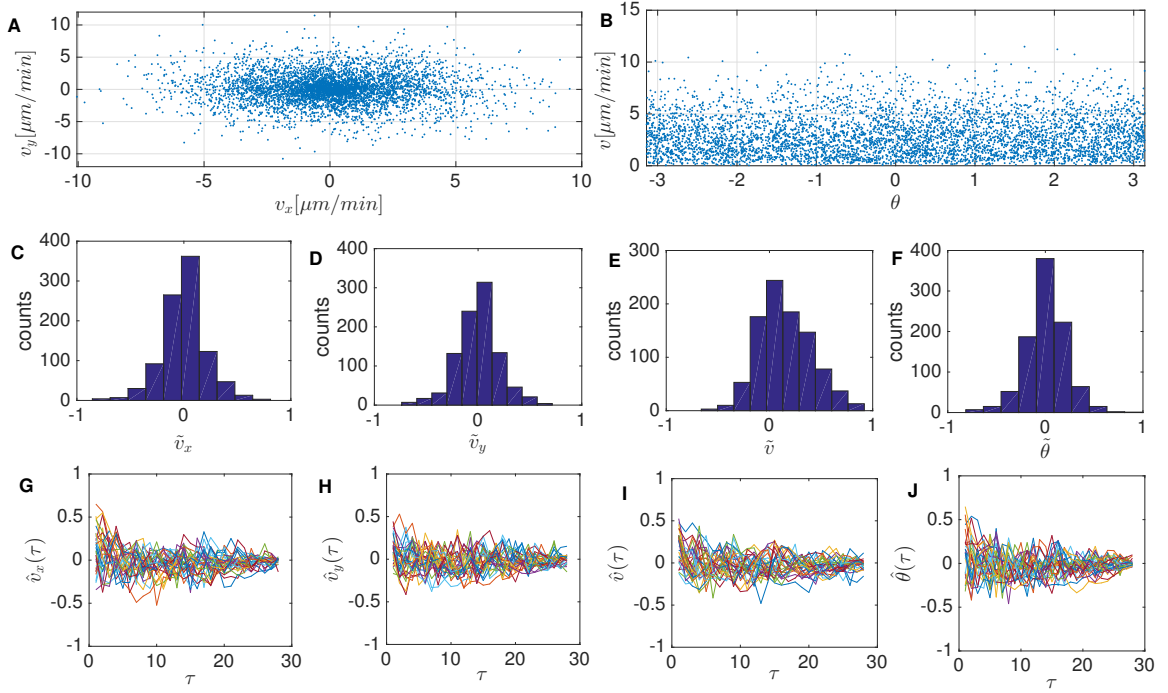

**Figure S6: Statistics of motional data for zone 1 in the dendritic-cell experiment.** First row: scatter plot for the velocity components $v_y$ vs. $v_x$ (**A**) and for the velocity norm $v$ vs. the directional angle $\theta$ (**B**). Second row: histograms for the pairwise correlation coefficient between the $x$-component of the velocities (**C**), between the $y$-component of the velocities (**D**), between the velocity magnitude (**E**), and between the angles (**F**). Third row: autocorrelation function for the $x$-component of the velocity (**G**), the $y$-component of the velocity (h), the magnitude of the velocity (**I**), and the angle (**J**). The outcomes of all cells (depicted in different colors) are all shown to give a visual representation of the variability of the autocorrelation. For the sake of clarity in panels (**A**), (**B**) (**G**)-(**J**), the data shown in this figure correspond to the low-density case, however, the high-density case yields analogous results.

components $v_y$ versus $v_x$ in Fig. S8A. Also, it is clear from the scatter plots $v$ versus $\theta$ (panel $b$) that the longest steps are typically those along the positive horizontal direction ($\theta = 0$). The pairwise correlations (panels C-F) exhibit distributions that are less symmetric than those in zone 1: in particular, $\tilde{v}_x$ and $\tilde{v}$ display a positive skewness and, overall, the indices of central position turn out to be $\overline{\tilde{v}_x} = 0.107$, $\text{mdn}(\tilde{v}_x) = 0.078$, $\text{std}(\tilde{v}_x) = 0.260$; $\overline{\tilde{v}_y} = 0.007$, $\text{mdn}(\tilde{v}_y) = 0.009$, $\text{std}(\tilde{v}_y) = 0.270$; $\overline{\tilde{v}} = 0.217$, $\text{mdn}(\tilde{v}) = 0.208$, $\text{std}(\tilde{v}) = 0.345$; $\overline{\tilde{\theta}} = 0.004$, $\text{mdn}(\tilde{\theta}) = 0.008$, $\text{std}(\tilde{\theta}) = 0.264$. Further, the auto-covariance functions for the velocity components, the velocity magnitude and the direction angle are still fluctuating around zero with poor dependence on time (panels G-J).

These results are robust with respect to cell density, and suggest that the picture of the isotropic random walk is no longer suitable to describe the motion in zone 2.

We now move to the analysis of the velocity distributions. We preliminary obtain $\overline{v_x} = 0.898$, $\text{mdn}(v_x) = 0.550$, $\text{std}(v_x) = 2.066$; $\overline{v_y} = -0.053$, $\text{mdn}(v_y) = -0.013$, $\text{std}(v_y) = 2.031$; $\overline{v} = 2.390$, $\text{mdn}(v) = 1.922$, $\text{std}(v) = 1.868$. The distributions of velocities and of angles in Fig. S9 show that, while $v_y$ is still exponentially distributed with an average comparable to the one obtained in the free environment (zone 1), $v_x$ is no longer symmetric around zero. Indeed, the distribution of $v_x$ is still exponential for negative $v_x$, while it appears to be half normal for positive $v_x$. More precisely, the density histogram for $v_x > 0$ is fitted by $\mathcal{N}_{1/2}(x;\sigma) = \sqrt{2/(\pi\sigma^2)}e^{-x^2/(2\sigma^2)}$ with best-fit coefficient $\sigma = 2.48 \pm 0.03$. This implies a broader dispersion for step lengths along the $x$-direction with large, positive steps getting more likely. Moreover, the distribution of the magnitude of the velocity is now best-fitted by a half-normal $\mathcal{N}_{1/2}(x;\sigma)$ with best-fit coefficient $\sigma = 3.0 \pm 0.1$, see Fig. S9E. Thus, even in zone 2 and unlike the wound-healing experiment, the distributions for $v_x$, $v_y$ and $v$ are well-behaved

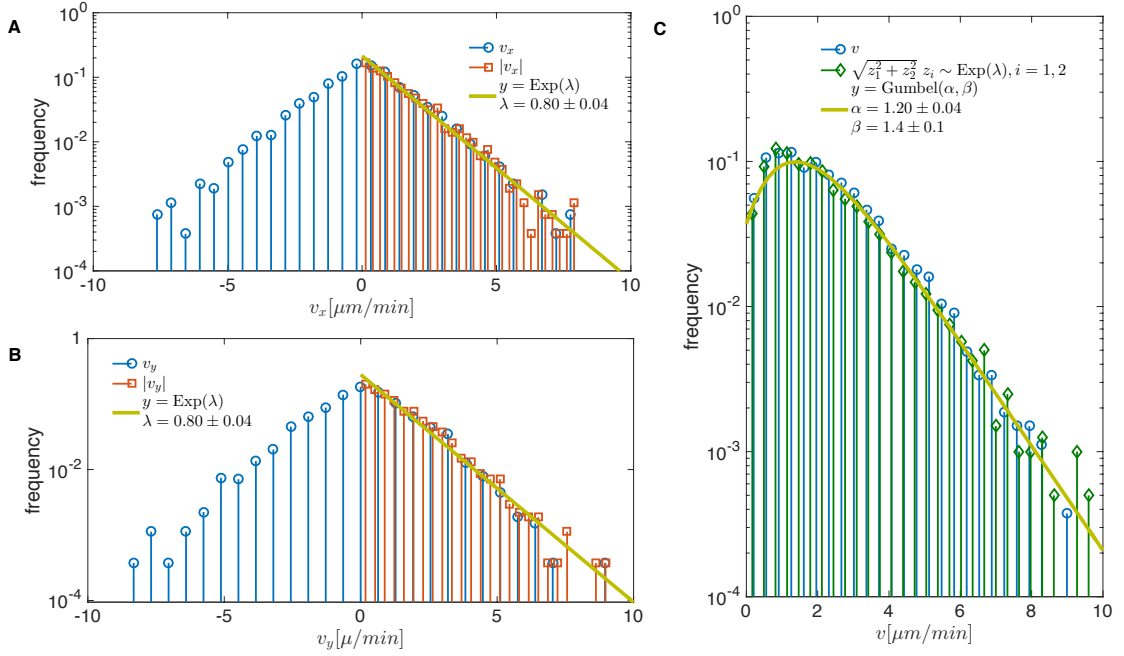

**Figure S7: Empirical distributions for cell velocities for zone 1 in the dendritic-cell experiment.** Density histogram for the $x$- and $y$-component of the velocity, panels (**A**) and (**B**), respectively, and for the magnitude $v$ (**C**) in a semi-logarithmic scale plot, with the related fits (black solid line). To highlight the symmetry of the distributions of $v_x$ and $v_y$, in panels (**A**) and (**B**) we show that the histogram for their absolute values $|v_x|$ and $|v_y|$ is nicely overlapped with the positive branch of the original histogram. Also, in order to corroborate the claim of an exponential distribution with mean $1/0.40 = 2.5$, in panel (**C**) we also show the histogram for the magnitude $z$ of a two-dimensional vector **z** whose components $z_1, z_2$ are independent and identically distributed from the same exponential distribution $\mathrm{Exp}(0.40)$: remarkably, the density histograms of $z$ and $v$ are nicely overlapped and well fitted by a Gumbel distribution.

distributions with short tails. This indicates that in the wound-healing experiment relatively large fluctuations are present.

## S6    Estimate of positional uncertainty

In this Section, we describe how we estimated the uncertainty in the cell positions resulting from a finite pixel size.

For either experiment, we chose the grayscale image of a single, representative cell, see Fig. S10A. We then resampled the image as follows. For each pixel we consider its color intensity, $C$, and divide the pixel into four equal subpixels. We draw randomly the intensities $c_i$, $i = 1, \cdots, 4$ in each subpixel according to a multinomial distribution with event probability $p_i = 1/4$, and a total number of trials equal to $4C$. As a result, the average of the intensities across the subpixels will equal the intensity of the original pixel, i.e., $1/4 \sum_{i=1}^{4} c_i = C$. By repeating this operation for all pixels in the original image, we produce a resampled image, for which we compute the cell center **r**, see Fig. S10B.

By repeating the procedure above, we obtain a population of resampled images, and the fluctuations of **r** across these samples reflect the uncertainty resulting from the finiteness of pixel size in the original image. As a result, we

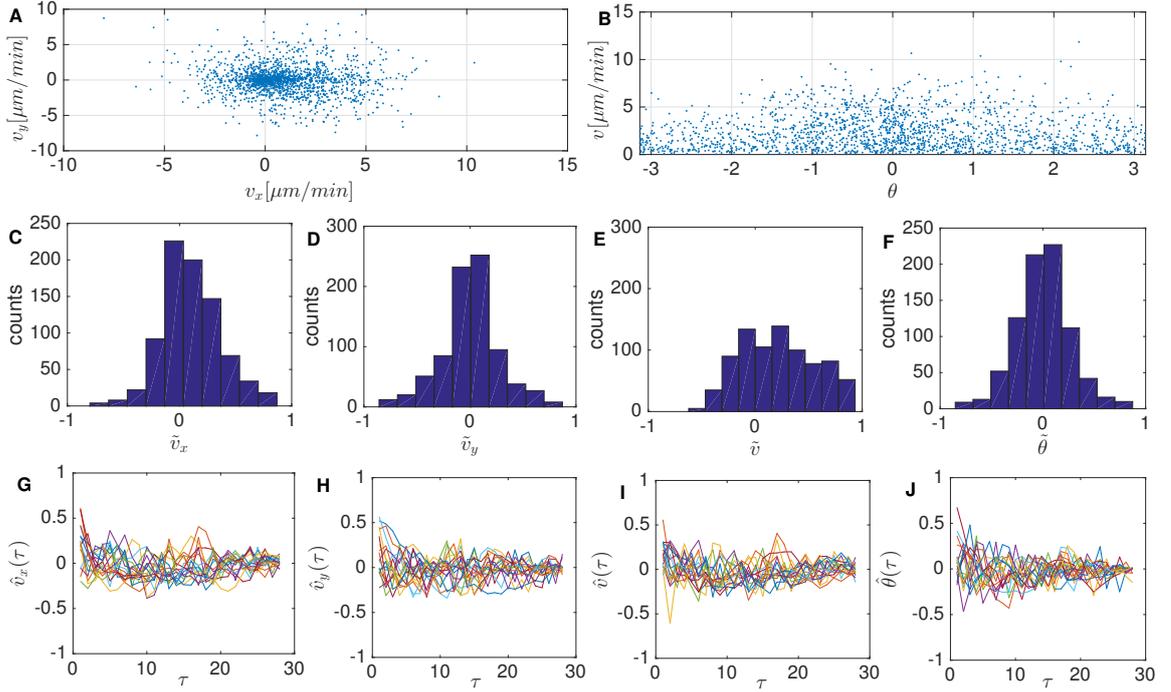

**Figure S8: Statistics of motional data for zone 2 in the dendritic-cell experiment.** First row: scatter plot for the velocity components $v_y$ vs. $v_x$ (**A**) and for the velocity norm $v$ vs. the directional angle $\theta$ (**B**). Second row: histograms for the pairwise correlation coefficient between the $x$-component of the velocities (**C**), between the $y$-component of the velocities (**D**), between the velocity magnitude (**E**), and between the angles (**F**). Third row: autocorrelation function for the $x$-component of the velocity (**G**), the $y$-component of the velocity (**H**), the magnitude of the velocity (**I**), and the angle (**J**). The outcomes of all cells (depicted in different colors) are all shown to give a visual representation of the variability of the autocorrelation. For the sake of clarity in panels (**A**) and (**B**), the data shown corresponds in this figure to the low-density case, however, the high-density case yields analogous results.

interpret the standard deviation $\sigma$ of each component of $\mathbf{r}$ across all samples as a rough estimate of the uncertainty on the cell position resulting from the finite pixel size.

For both components, we obtain $\sigma \sim 0.15\,\mu\text{m}$ and $\sigma \sim 0.07\,\mu\text{m}$ for the cancer-cell and wound-healing experiment, respectively.

The nominal cell positions are then obtained by rounding off the cell coordinates to $\sigma$:

$$\boldsymbol{r}_i(t) \to \sigma \left\lfloor \frac{\boldsymbol{r}_i(t)}{\sigma} \right\rfloor, \tag{S42}$$

in such a way that all positions that differ by less than $\sigma$ are assigned the same nominal position, see Fig. 1.

The cell positions obtained with (S42) are then used to compute the directions of motion according to the velocity definition in Section 2.1 and Eq. (7), where some of these directions are, in general, not defined because of the coarsening scheme described in (S42). The resulting empirical averages are obtained with Eqs. (1), (8) and (9): for example, the average correlation reads

$$\langle C(\mathbf{S}) \rangle_{\text{ex}} = \frac{1}{N_p T} \sum_{i<j=1}^{N} \sum_{t=1}^{T} \mathbf{s}_i(t) \cdot \mathbf{s}_j(t). \tag{S43}$$

We then let all ill-defined directions of motion $\mathbf{s}_i(t)$ which enter in the right-hand side of Eq. (S43) vary between 0 and $2\pi$, and compute the related interval in which $\langle C(\mathbf{S}) \rangle_{\text{ex}}$ varies. The lower and upper bounds of this interval

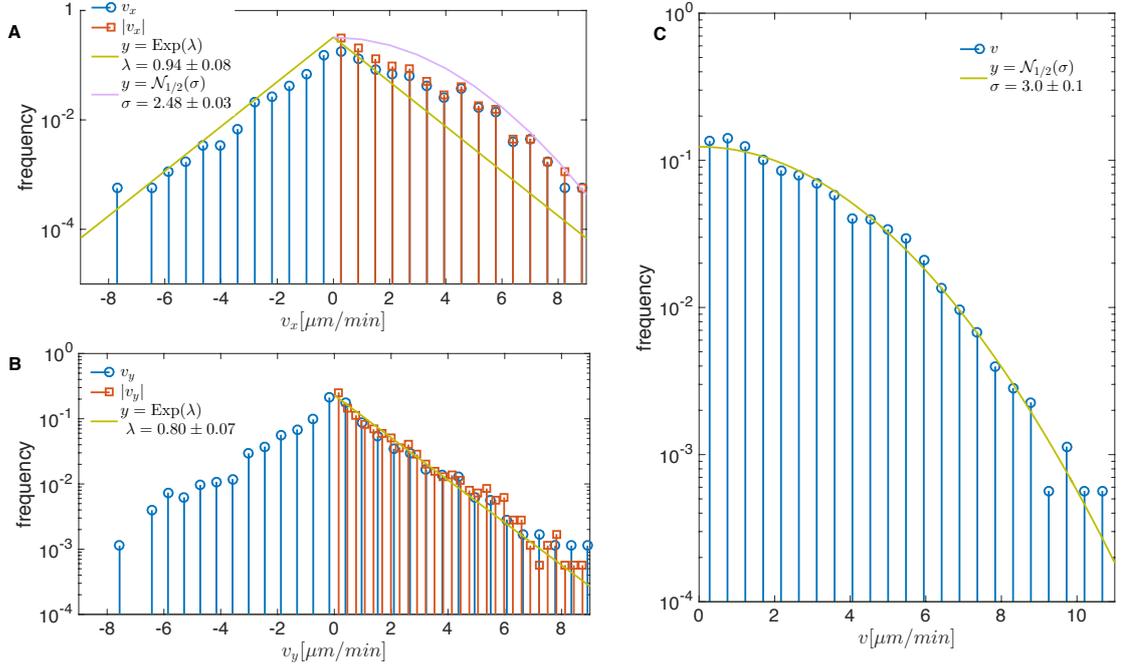

**Figure S9: Empirical distributions for cell velocities for zone 2 in the dendritic-cell experiment.** Density histogram for the $x$- and $y$-component of the velocity, panels (**A**) and (**B**), respectively, and for the magnitude $v$ (**C**) in a semi-logarithmic scale plot, with the related fits (solid black line). To highlight the asymmetry and the symmetry of the distributions of, respectively, $v_x$ and $v_y$, in panels (**A**) and (**B**) we overlap the histogram for their absolute values $|v_x|$ and $|v_y|$ to the positive branch of the original histogram. The comparison is fine only for $v_y$ which follows an exponential distribution quantitatively consistent with the one obtained for zone 1. As for $v_x$, the exponential distribution works only for the negative branch, with a mean value slightly smaller than the one obtained in zone 1, while for the positive branch the best fit is given by a half-normal distribution. A half-normal distribution is also found to best-fit the distribution for the velocity magnitude.

are the values $C_{\text{ex}}^{\min(\max)}$ that we use in the MEb method, cf. Eq. (S18). The same procedure is used to compute the bounds for each component of the polarization $\boldsymbol{M}$. The resulting numerical values for the lower and upper bounds for the wound-healing and dendritic-cell experiment are shown in Tables S1 and S2, respectively.

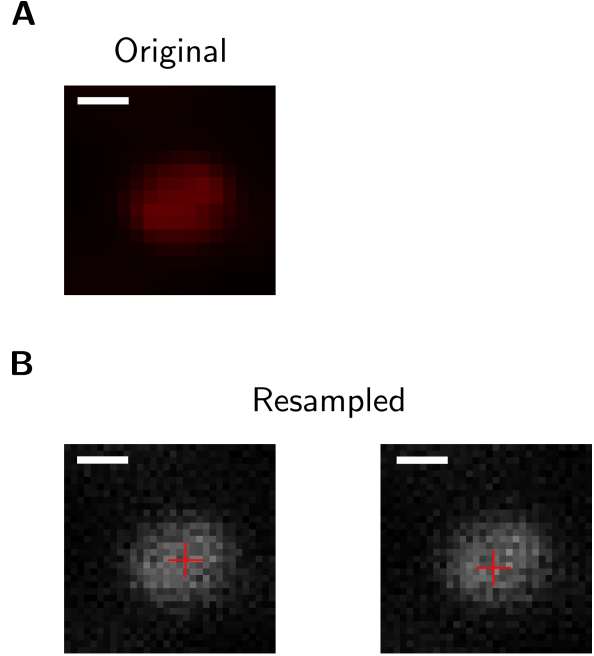

**Figure S10: Estimate of the positional error.** (**A**) Original image of a representative cell in the wound-healing experiment, cf. Fig. 3A. (**B**) Two representative images obtained by resampling the image in (**A**), where for each image the cross indicates the cell center (scale bar 5 $\mu$m).

| | |
|---|---|
| $\langle C \rangle_{\min}$ | 0.16 |
| $\langle C \rangle_{\max}$ | 0.49 |
| $\langle \boldsymbol{M} \rangle_{\min}$ | (0.36 , -0.12) |
| $\langle \boldsymbol{M} \rangle_{\max}$ | (0.54 , 0.048) |

**Table S1: Confidence intervals for the empirical averages in the wound-healing experiment.** Lower and upper bounds for the empirical averages of the correlation $C$ and polarization $\boldsymbol{M}$.

| Density | High | | Low | |
|---|---|---|---|---|
| Zone | 1 | 2 | 1 | 2 |
| $\langle C \rangle_{\min}$ | -0.086 | -0.018 | -0.14 | -0.027 |
| $\langle C \rangle_{\max}$ | 0.15 | 0.27 | 0.14 | 0.28 |
| $\langle \boldsymbol{M} \rangle_{\min}$ | (-0.12 , 0.0025) | (0.22 , -0.13) | (0.036 , -0.098) | (0.27 , -0.043 ) |
| $\langle \boldsymbol{M} \rangle_{\max}$ | (0.0011 , 0.12) | (0.37 , 0.023) | (0.18 , 0.046) | (0.43 , 0.12) |

**Table S2: Confidence intervals for the empirical averages in the dendritic-cell experiment.** Results for the high- and low-density case and the high- and low-cytokine region, i.e., zones 1 and 2, respectively, are presented. For each experimental instance, we show the lower and upper bounds for the empirical averages of the correlation $C$ and polarization $\boldsymbol{M}$.



\end{document}